\documentclass[lettersize, journal]{IEEEtran}
\usepackage{array}
\usepackage{float}
\usepackage{caption}
\usepackage{color}

\usepackage{graphicx} \usepackage{float} 
\usepackage{subcaption}
\usepackage{cite}
\usepackage{amsmath,amssymb,amsfonts}
\usepackage{amsthm}
\usepackage{graphicx}
\usepackage{textcomp}
\usepackage{xcolor}
\usepackage{svg}
\usepackage{algorithm}
\usepackage{algorithmic}
\usepackage{mathtools}
\usepackage{lipsum}

\newtheorem{definition}{Definition}

\newtheorem{theorem}{Theorem}

\newtheorem{lemma}{Lemma}

\newtheorem{corollary}{Corollary}

\newcounter{TempEqCnt}

\allowdisplaybreaks
\makeatletter
\def\ScaleIfNeeded{%
\ifdim\Gin@nat@ width>\linewidth \linewidth \else \Gin@nat@width
\fi } \makeatother

\graphicspath{{Fig/}}
\begin{document}
\title{Resource Allocation for Pinching-Antenna Systems (PASS)-enabled NOMA Communications}
\author{Songtao Xue, Jingjing Zhao, Kaiquan Cai, Xidong Mu, Zhenyu Xiao, and Yuanwei Liu,~\IEEEmembership{Fellow,~IEEE}

   \thanks{S. Xue, J. Zhao, K. Cai, and Z. Xiao are with the School of Electronics and Information Engineering, Beihang University, Beijing, China, and also with the State Key Laboratory of CNS/ATM, Beijing, China. (e-mail:\{xuesongtao, jingjingzhao, caikq, xiaozy\}@buaa.edu.cn). }
    \thanks{X. Mu is with the Centre for Wireless Innovation (CWI), Queen's University Belfast, Belfast, BT3 9DT, U.K. (e-mail: x.mu@qub.ac.uk).}
    \thanks{Y. Liu is with the Department of Electrical and Electronic Engineering, the University of Hong Kong, Hong Kong, China (e-mail: yuanwei@hku.hk).}
 }

\maketitle
\begin{abstract}
Pinching-antenna systems (PASS) have emerged as a promising technology due to their ability to dynamically reconfigure wireless propagation environments. A novel PASS-based multi-user non-orthogonal multiple access (NOMA) framework is proposed by exploiting the waveguide-division (WD) transmission characteristic. Specifically, each NOMA user cluster is served by one dedicated waveguide, and the corresponding pinching beamforming is exploited to enhance the intra-cluster performance while mitigating the inter-cluster interference. Based on this framework, a 
sum-rate maximization problem is formulated for jointly optimizing power allocation, pinching beamforming, and user scheduling. To solve this problem, a two-step algorithm is developed, which decomposes the original problem into two subproblems. For the joint power allocation and pinching beamforming design, a penalty dual decomposition (PDD) algorithm is proposed to obtain the locally optimal solutions. Specifically, the coupling constraints are alleviated through augmented Lagrangian relaxation, and the resulting augmented Lagrangian (AL) problem is decomposed into four subproblems, which are solved by the block coordinate descent (BCD) method.  For the user scheduling, a low-complexity matching algorithm is developed to solve the user-to-waveguide assignment problem. Simulation results demonstrate that 1) the proposed PASS-based NOMA framework under the WD transmission structure achieves significant sum-rate gain over conventional fixed-position antenna systems and orthogonal multiple access (OMA) scheme; and 2) the proposed matching-based user scheduling algorithm
 achieves near-optimal user-waveguide association with low computational complexity.
\end{abstract}
\section{Introduction}

In recent years, flexible-antenna systems have gained significant research attention in wireless communications due to their ability to reconfigure the wireless propagation environment~\cite{flexible_antenna1,flexible_antenna2}. For instance, novel antenna architectures such as fluid-antenna (FA) and movable-antenna (MA) systems can dynamically adjust the positions of antennas to enhance channel conditions~\cite{FA1,Movable1,Movablexzy1}. However, a critical limitation of these systems is that antennas movement is always constrained within a few wavelengths~\cite{Movablexzy2}. While the movement in a confined region
 is effective for mitigating small-scale fading, it is insufficient for combating large-scale path loss. Another example of flexible-antenna systems is the reconfigurable intelligent surface (RIS), which reconfigures the propagation environment by intelligently adjusting phase shifts~\cite{RIS}. Although RISs can be leveraged to provide additional virtual links for communication users, they often suffer from the double-fading issue, which also limits their capability to mitigate large-scale path loss. Consequently, how to combat large-scale path loss remains a key challenge for existing flexible-antenna technologies~\cite{Movable3, Movable4}.

To tackle this challenge, pinching antenna systems (PASS) have recently emerged as a novel flexible-antenna technique~\cite{pinching0,xuxiaoxia}. PASS employ dielectric waveguides with low propagation losses to transmit signals, where pinching dielectric particles known as pinching antennas (PAs) are deployed at any position along the waveguide
for radiating signals into the free space. By dynamically deploying these PAs, the system can direct radiation to establish strong line-of-sight (LoS) links with communication users, and significantly reduce large-scale path loss. This capability results in significant gains in received signal strength and coverage reliability. Moreover, PASS exhibit excellent scalability compared to conventional antenna systems, because adding or removing PAs at selected positions on the waveguide is easy to implement and does not change the internal structure of the PASS.

One characteristic of PASS is that multiple PAs activated on a single dielectric waveguide transmits the same signal~\cite{pinching1}. Considering this limitation, when the number of users to be served simultaneously exceeds the number of waveguides, conventional spatial multiplexing methods cannot meet the service requirements, which motivates the use of non-orthogonal multiple access (NOMA). Specifically, NOMA exploits superposition coding at transmitter and successive interference cancellation (SIC) at receiver to serve multiple users in the time, frequency, or code domain. This principle of NOMA inherently aligns with the requirement that all PAs on a single waveguide transmit the same signal in PASS. Therefore, the combination of NOMA and PASS in the multi-user scenario deserves further investigation.
\subsection{Related Works}
With the growing importance of the aforementioned benefits, extensive research works
have been devoted to PASS~\cite{pinching-arraygain,pinching-ofdma,pinching-related1,pinching-related2,pinching-related3,pinching-related4}. To explore the performance of PASS, the achievable array gain was analyzed in~\cite{pinching-arraygain}, which demonstrated the existence of an optimal inter-antenna spacing. Furthermore, the authors of~\cite{pinching-ofdma} proposed an orthogonal frequency-division multiple access (OFDMA) framework and developed a low-complexity two-stage resource allocation algorithm to overcome the severe inter-symbol interference in multi-user PASS. Since the effective activation of PAs at appropriate locations is a critical issue in PASS, some research has focused on the design of pinching beamforming. To maximize the downlink transmission rate, the authors of~\cite{pinching-related1} considered a single-user system and developed a two-stage algorithm by jointly optimizing transmit and pinching beamforming design. The authors of~\cite{pinching-related2} integrated a multi-user multiple-input single-output (MISO) multicast scenario and proposed a unified cross-entropy optimization framework to jointly optimize PA placement and transmit beamforming. Moreover, a graph neural network was employed in~\cite{pinching-related4} to jointly optimize the PA placement and power allocation by modeling the downlink PASS as a bipartite graph. For the uplink scenario, the authors of~\cite{pinching-related3} developed a decoupled optimization approach to maximize the minimum user data rate, which uniquely transformed the antenna positioning into a convex problem and provided a closed-form solution for resource allocation.

 To facilitate multi-user communications, the combination of NOMA and PASS has attracted some research contributions recently. In~\cite{related-ding}, the authors proposed a low-complexity placement design for pinching-antenna-assisted networks, where a closed-form solution was derived to optimize the placement of multiple PAs in both time division multiple access (TDMA) and NOMA schemes. Furthermore, the authors of~\cite{related-wkd} formulated a sum rate maximization problem with PA activation and proposed a low-complexity matching algorithm to obtain a near-optimal solution. In ~\cite{related-analytical}, the authors conducted a rigorous analytical optimization for a simplified two-user NOMA framework, deriving a closed-form expression of PA location that ensured user fairness. To minimize power consumption, the authors of~\cite{related-rq} studied a two-user NOMA system, developing a two-stage algorithm for PA positioning and providing a comprehensive performance comparison against the orthogonal multiple access (OMA) scheme. Considering the uplink PASS framework, the authors of~\cite{related-zm} proposed an alternating optimization (AO) framework in which a modified particle swarm optimization (PSO) algorithm was utilized to optimize the antenna position while a low-complexity solution was derived for resource allocation. In addition, the authors of~\cite{related-xyq} developed a two-user cognitive radio-inspired NOMA framework, and proposed a joint PA positions and power allocation design to maximize the secondary user's data rate under the quality-of-service (QoS) constraint imposed by the primary user. Moreover, a PASS-aided NOMA protocol was investigated in~\cite{related-lyx} within a wireless powered communication network, where resource allocation and PAs positioning are jointly optimized for maximizing the sum rate of the uplink system.


\subsection{Motivations and Contributions}
While the combination of NOMA and PASS has been explored, existing research has primarily focused on simple scenarios, typically involving a single waveguide serving a single group of NOMA users. To explore the full potential of this technology in large-scale networks, it is essential to consider more complex deployments with multiple waveguides. However, the multi-waveguide architecture requires the complex joint optimization of transmit beamforming and pinching beamforming, which inevitably introduces significant computational challenges.
To address this issue, this paper proposes a novel PASS-NOMA framework based on a waveguide-division (WD) architecture. For the WD structure, each waveguide is responsible for transmitting a single data stream dedicated to its intended user, enabling the baseband to carry out simple power allocation and significantly reducing design complexity. Furthermore, the corresponding pinching beamforming within the dedicated waveguide is exploited to enhance intra-cluster performance while effectively mitigating inter-cluster interference.


However, the NOMA system based on the WD architecture also introduces certain new challenges. On the one hand, a critical user scheduling problem arises from this architecture. This issue involves two coupled decisions: the partitioning of users into distinct NOMA clusters, and the assignment of these clusters to the available waveguides. Due to the vast number of possible assignments, finding the optimal schedule via conventional exhaustive search becomes intractable. On the other hand, for a given user scheduling scheme, the joint optimization of power allocation and pinching beamforming remains a highly complex and non-convex problem due to the complex exponential and fractional terms and highly coupled optimization variables. In addition, unlike the single-waveguide PASS architecture, the multi-waveguide system must address the challenge of mitigating inter-cluster interference by leveraging pinching beamforming. Motivated by these challenges, this paper develops a novel WD-PASS-based NOMA framework. The contributions of this paper are summarized as follows:
\begin{itemize}
\item We consider a novel WD-PASS-based NOMA communications framework, where each user cluster is served by a dedicated waveguide. Based on this framework, we formulate a joint power allocation, pinching beamforming, and user scheduling optimization problem to maximize the system's sum rate, while satisfying the base station (BS) power constraint and user QoS limitation. 

\item We decompose the original formulation into two problems. For the joint power allocation and pinching beamforming design, we transform the non-convex optimization problem into a weighted minimum mean square error (WMMSE) problem. Then, we develop a penalty dual decomposition (PDD) algorithm to alleviate the stringent coupling constraints by using Lagrangian relaxation, where the block coordinate descent (BCD) method is employed to alternatively optimize each block of variables. For the user scheduling problem, we invoke a low-complexity matching-based algorithm to obtain the near-optimal user-waveguide association
solution. 

\item Our numerical results demonstrate that 1) the proposed  WD-PASS-based NOMA  framework achieves a significant sum-rate gain over the conventional antenna systems and the OMA scheme, and 2) the proposed low-complexity matching-based algorithm achieves near-optimal solution, which has similar performance to the exhaustive search scheme.
\end{itemize}
\subsection{Organization and Notations} 
The remainder of this paper is structured as follows.
Section II introduces the system model for WD-PASS-based NOMA scenario and formulates an optimization problem. Section III proposes the joint power allocation, pinching beamforming, and user scheduling optimization algorithm. In Section V, simulations are provided to verify the effectiveness and superiority of our proposed WD-PASS-based NOMA framework. Section VI concludes the article.

\textit{Notations:}
Scalars, vectors and matrices, are denoted by lowercase letters, boldface lowercase letters, and boldface capital letters, respectively. $\mathbb{C}^{M\times N}$ denotes the complex matrices with the space $M\times N$. $\mathbf{a}^T$, $\mathbf{a}^H$, and $\mathbf{a}^*$ represent the transpose, conjugate transpose, and conjugate of the vector $\mathbf{a}$, respectively. $\text{Re}(\mathbf{A})$ and $\text{Im}(\mathbf{A})$ denote the real and imaginary part of matrix $\mathbf{A}$. $\mathcal{CN}\left(m,n\right)$  denotes a Gaussian distribution 
with mean $m$ and variance $n$. $\|\mathbf{a}\|$ represents the norm of the vector $\mathbf{a}$. $|\cdot|$ is the absolute value of a scalar. $\mathbf{1}_{M\times N}$ denotes the $M \times N$ all-ones matrix. 
\section{System Model and Problem Formulation}
\begin{figure}
    \centering
\includegraphics[width=1.02\linewidth]{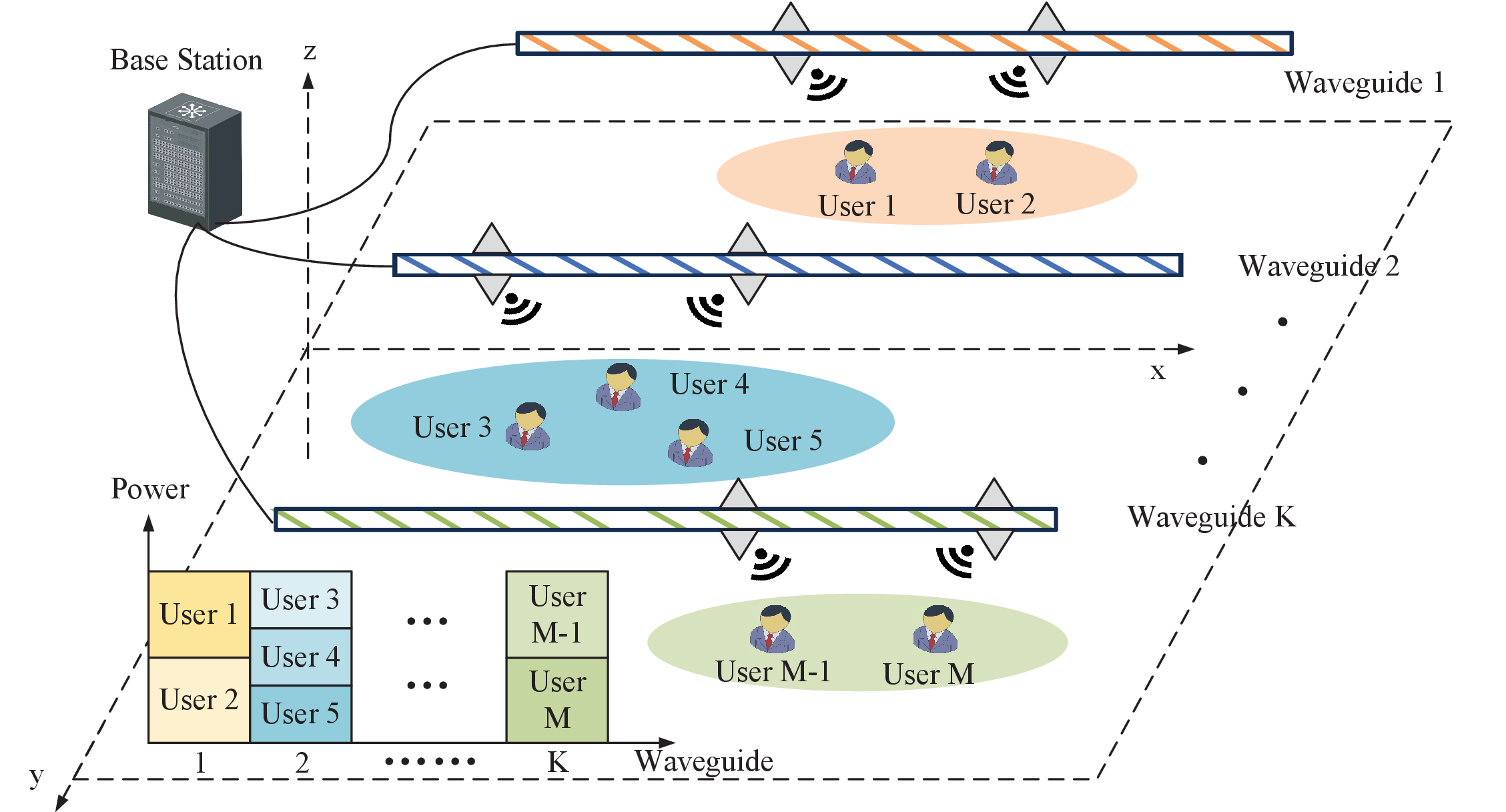}
    \caption{System model for WD-PASS-based NOMA scenario.}
    \label{fig:sys_model}
\end{figure}
As shown in Fig.~\ref{fig:sys_model}, we consider a downlink communication scenario where a BS consisting of $K$ waveguides serves $M$ single-antenna users in different clusters. Denote $\mathcal{K}=\{1,..., K\}$ and $\mathcal{D}=\{\mathcal{D}_1,\mathcal{D}_2,...,\mathcal{D}_K\}$ as the waveguide set and the user cluster set, respectively, where $\mathcal{D}_k$ represents the user cluster served by the $k$-th waveguide. For each waveguide, $N$ PAs are activated to jointly serve a cluster of users via the NOMA protocol based on the power-domain superposition. Each waveguide spans a length of $L$, while the spacing between two waveguides is denoted by $W$. We set a three-dimensional (3D) coordinate system, where the origin is located at the position of the BS. Specifically, the waveguide is parallel to the $x$-axis at a height $d$, and users are distributed within the $x$-$y$ plane. The position of the $m$-th user is given by $\boldsymbol{\psi}^{\mathrm{u}}_{m}=\left[x^u_{m},y^u_{m},0\right]^T$. Then, the position of the $n$-th PA on the $k$-th waveguide is given by $\boldsymbol{\psi}^{\mathrm{p}}_{k,n}=\left[x^{\mathrm{p}}_{k,n},y^{\mathrm{p}}_{k,n},d\right]^T$. To avoid antenna coupling, the distance between any two adjacent PAs is restricted to be not smaller than $\Delta$, which is normally set as half of the wavelength. The position of the feed point on the $k$-th waveguide is denoted as $\boldsymbol{\psi}^{{\text{feed}}}_k=\left[0, y^{\mathrm{p}}_{k,n},d\right]$. Denote $\mathcal{C}= \{c_m^k\in\{0,1\}, m\in\mathcal{M},k\in\mathcal{K}\}$ as the set of users scheduling index. Specifically, $c_m^k=1$ if the $m$-th user is served by the $k$-th waveguide, and $c_m^k=0$ otherwise.  Let $p_{m}$ represent the transmit power allocated to $m$-th user with $\sum_{m=1}^Mp_{m}= P_\text{max}$, where $P_\text{max}$ denotes the maximum BS transmit power. 

\subsection{Channel Model}
 The channel vector between PAs on the $k$-th waveguide and the $m$-th user is denoted by

\begin{align}
    & \mathbf{h}_{k,m}\left(\mathbf{x}^{\text{p}}_{k}\right)\nonumber\\
    &= \left[\frac{\eta e^{-j\frac{2\pi}{\lambda}{\left\|\boldsymbol{\psi}_{m}^{\text{u}}-\boldsymbol{\psi}_{k,1}^{\text{p}}\right\|}}}{\left\|\boldsymbol{\psi}_{m}^{\text{u}}-\boldsymbol{\psi}_{k,1}^{\text{p}}\right\|}, \cdots, \frac{\eta e^{-j\frac{2\pi}{\lambda}{\left\|\boldsymbol{\psi}_{m}^{\text{u}}-\boldsymbol{\psi}_{k,N}^{\text{p}}\right\|}}}{\left\|\boldsymbol{\psi}_{m}^{\text{u}}-\boldsymbol{\psi}_{k,N}^{\text{p}}\right\|}\right]^T,
\end{align}
where $m\in\mathcal{M}$, $k\in\mathcal{K}$, $\eta =\frac{\lambda}{4\pi}$, $\mathbf{x}^{\text{p}}_{k}=\left[x^\mathrm{p}_{k,1},\cdots,x^{\mathrm{p}}_{k,N}\right]^T$ is the $x$-axis positions vector for PAs on the $k$-th waveguide, and $\lambda$ denotes the signal wavelength. The distance between the $m$-th user and the $n$-th PA on the $k$-th waveguide is given by
\begin{equation}
\begin{aligned}
\label{distance_u_pa}
&\left\|\boldsymbol{\psi}_{m}^{\text{u}}-\boldsymbol{\psi}^{\text{p}}_{k,n}\right\|=\sqrt{\left(x_{m}^{\text{u}}-x^{\text{p}}_{k,n}\right)^2 +\left(y_{m}^{\text{u}}-y^{\text{p}}_{k,n}\right)^2+d^2},
\end{aligned}
\end{equation}
where $\forall m\in \mathcal{M}, k,k'\in \mathcal{K}$. The signal propagation response in the waveguide is denoted as
 \begin{equation}   
\label{eq:channel}   \mathbf{g}_{k}\left(\mathbf{x}^{\text{p}}_{k}\right)=\left[e^{-\frac{2\pi}{\lambda_g}\|\boldsymbol{\psi}^{\mathrm{p}}_{k,1}-\boldsymbol{\psi}^{\text{feed}}_{k}\|},\cdots,e^{-\frac{2\pi}{\lambda_g}\|\boldsymbol{\psi}^{\mathrm{p}}_{k,N}-\boldsymbol{\psi}^{\text{feed}}_{k}\|}\right]^T,
\end{equation}
where $\lambda_{\text{g}}=\frac{\lambda}{n_{\text{eff}}}$ is the guided wavelength with $n_{\text{eff}}$ representing the effective refractive index of a dielectric waveguide. Since PAs on each waveguide are distributed along the $x$-axis direction, the distance between the $n$-th PA and the feed point on the $k$-th waveguide is $x_{k,n}$.

\subsection{Signal Model}
Let $s_m$ represent the signal transmitted by
the $k$-th waveguide to the $m$-th user with $\mathbb{E}\left[s_ms_m^*\right]=1$. The received signal at the $m$-th user served by the $k$-th waveguide is then given by
\begin{equation}
\begin{aligned}
\label{eq:recieved_signal}
        y_{k,m} &= c_m^k\sqrt{p_m}\mathbf{h}^T_{k,m}\left(\mathbf{x}^{\text{p}}_{k}\right)\mathbf{g}_{k}\left(\mathbf{x}^{\text{p}}_{k}\right)s_m \\
    &+\sum_{m'\neq m}^M c_{m'}^k\sqrt{p_{m'}}\mathbf{h}^T_{k,m}\left(\mathbf{x}^{\text{p}}_{k}\right)\mathbf{g}_{k}\left(\mathbf{x}^{\text{p}}_{k}\right)s_{m'}
    \\
    & +\sum_{k'\neq k}^K\sum_{i= 1}^M c_{i}^{k'}\sqrt{p_{i}}\mathbf{h}^T_{k',m}\left(\mathbf{x}^{\text{p}}_{k'}\right)\mathbf{g}_{k'}\left(\mathbf{x}^{\text{p}}_{k'}\right)s_{i} + n_m,
\end{aligned}
\end{equation}
where the first, second, and third terms on the right-hand side (RHS) of the equation denote the expected signal, intra-cluster interference, and inter-cluster interference, respectively, and $n_m\sim \mathcal{CN}(0, \sigma^2)$ represents the additive white Gaussian noise (AWGN) at the $m$-th user. 

In the multi-user downlink NOMA system, the user with a stronger channel condition can apply SIC to decode the signal of the user who has a weaker channel condition before decoding their own signal. Since the channel gains can be modified by PAs positions, the optimal decoding order can be any of the possible decoding orders in each cluster. Therefore, an exhaustive search is needed over all the decoding orders within each user cluster. Let $\pi_k\left(m\right)$ denote the decoding order for user $m$ in cluster $\mathcal{D}_k$. For any two users $l$ and $m$ served by the $k$-th waveguide satisfying $\pi_k(m)\geq\pi_k(l)$, the signal-to-interference-plus-noise ratio (SINR) of the $m$-th user to decode the signal of user $l$ is expressed as
\begin{equation}
\label{eq:SINR}
\gamma_{m\rightarrow{}l}^{k} =
\frac{c_{l}^k{p_{l}}|\mathbf{h}_{k,m}^T\left(\mathbf{x}^{\text{p}}_{k}\right)\mathbf{g}_k\left(\mathbf{x}^{\text{p}}_{k}\right)|^2}{I_{\text{intra}} + I_{\text{inter}} + \sigma^2},
\end{equation}
where the intra-cluster and inter-cluster user interference can be respectively expressed as
\begin{equation}
\begin{aligned}
    \label{interference}
    &I_{\text{intra}} = \sum_{l' \in D_k, \pi_k(l') > \pi_k(l)}c_{l'}^k{p_{l'}}|\mathbf{h}_{k,m}^T\left(\mathbf{x}_{\text{p}}\right)\mathbf{g}_k\left(\mathbf{x}^{\text{p}}_{k}\right)|^2,
   \\& I_{\text{inter}}=\sum_{k'\neq k}^K\sum_{i \in D_{k'}} c_{i}^{k'}{p_{i}}|\mathbf{h}^T_{k',m}\left(\mathbf{x}^{\text{p}}_{k'}\right)\mathbf{g}_{k'}\left(\mathbf{x}^{\text{p}}_{k'}\right)|^2.
\end{aligned}
\end{equation}
The rate for user $m$ to decode the signal of user $l$ is given by $R_{m\rightarrow{}l}^{k}=\log_2\left(1+\gamma_{m\rightarrow{}l}^{k}\right)$. Furthermore, the achievable rate at user $l$ to decode its own signal is $R_{l\rightarrow{}l}^{k}=\log_2\left(1+\gamma_{l\rightarrow{}l}^{k}\right)$. To perform the SIC successfully at the user $l$, the SIC decoding rate constraints  $R_{m\rightarrow{}l}^{k}\geq R_{l\rightarrow{}l}^{k}$ should 
be satisfied~\cite{SIC}. 


\subsection{Problem Formulation}
In this paper, we aim to maximize the sum rate of the downlink WD-PASS-based NOMA system, subject to the BS power constraint and user QoS limitation. The overall optimization problem can be formulated as follows:
\begin{subequations}
\label{eq:optimization_problem}
\begin{equation}
\label{eq:objective_function}    \max_{\mathcal{C},\mathbf{P},\mathbf{X}}\sum_{k=1}^K\sum_{m=1}^M{R}^{k}_{m\rightarrow{}m},
\end{equation}
\begin{equation}
\label{eq:constrait_x}
   {\rm{s.t.}} \  \ 0\leq x_{k,n}^{\text{p}} \leq L, \forall n\in \mathcal{N},\forall k\in \mathcal{K},
\end{equation}
\begin{equation}
\label{eq:constraint_Delta}
  x_{k,n+1}^{\text{p}}- x_{k,n}^{\text{p}} \leq \Delta, \forall 1\leq n<N,\forall k\in \mathcal{K},
\end{equation}
\begin{equation}
\label{eq:constraint_p}
 \sum_{m=1}^{M}{p}_{m}\leq {P}_{\text{max}},
\end{equation}
\begin{equation}
\label{eq:constraint_qos}
 R_{m\rightarrow{}m}^{k}\geq R_{\text{min}}, \forall m\in \mathcal{D}_{k},\forall k\in \mathcal{K},
\end{equation}
\begin{equation}
\label{eq:constraint_sic}
 R_{m\rightarrow{}l}^{k}\geq R^{k}_{l\rightarrow{}l}, \forall m,l\in \mathcal{D}_{k},\forall k\in \mathcal{K},\pi(m)\geq\pi(l),
\end{equation}
\begin{equation}
\label{eq:constraint_num}
\sum_{m=1}^Mc_m^k\leq \bar{q},\forall k\in \mathcal{K},
\end{equation}
\begin{equation}
\label{eq:constraint_c}
\sum_{k=1}^Kc_m^k=1, \forall m\in \mathcal{M},
\end{equation}
\end{subequations}
where $\mathbf{P}=\{p_m|m\in \mathcal{M}\}$ and $\mathbf{X}=\{\mathbf{x}^{\text{p}}_{k},k\in \mathcal{K}\}$ represent the optimization variable sets of the power allocation coefficients and the PAs positions, respectively. $\eqref{eq:constrait_x}$ gives the movable region constraint of PAs along waveguides. $\eqref{eq:constraint_Delta}$ is the spacing constraint among PAs to avoid antenna coupling, where $\Delta$ denotes the minimum PA spacing. ~\eqref{eq:constraint_p} limits the transmit power not to exceed $P_\text{max}$ at BS. ~\eqref{eq:constraint_qos} is the QoS requirement of each user, where $R_{\text{min}}$ denotes the minimum data rate requirement. ~\eqref{eq:constraint_sic} guarantees the SIC perform successfully.~\eqref{eq:constraint_num} limits the number of users in each user cluster, where $\bar{q}$ denotes that the maximum number of users can be served by each waveguide. ~\eqref{eq:constraint_c} represents that each user can be served by one waveguide at maximum.

It is intractable to solve problem~\eqref{eq:optimization_problem} due to the following two main reasons. On the one hand, the user scheduling design involves two coupled tasks: the partitioning of users into distinct NOMA clusters, and mapping them to specific waveguides. Due to the vast number of possible assignments, it is challenging to obtain the optimal schedule by conventional exhaustive search methods. On the other hand, the joint optimization of power allocation and pinching beamforming remains a highly complex and non-convex problem due to the complex exponential expressions and the coupled optimization variables.

\section{Joint Power Allocation, Pinching Beamforming and User Scheduling Scheme}
 
In this section, we decouple the problem~\eqref{eq:optimization_problem} into the two subproblems. For power allocation and pinching beamforming design, we propose a PDD algorithm to address equality constraints by augmented Lagrangian relaxation. Further, the optimization problem is divided inito four subproblems to update each block of variables alternatively in a BCD manner. Then, we propose a matching algorithm for user scheduling with low computational complexity.

\subsection{PDD Algorithm for Power Allocation and Pinching Beamforming}
To begin with, for a given set of $\mathcal{C}$ and a fixed decoding order, the optimization problem~\eqref{eq:optimization_problem} can be simplified as follows:
\begin{subequations}
\label{eq:optimization_problem_xp}
\begin{equation}
\label{eq:objective_function_initial} \max_{\mathbf{P},\mathbf{X}}\sum_{k=1}^K\sum_{j=1}^{q_k}{R}^{k}_{j\rightarrow{}j},
\end{equation}
\begin{equation}
\label{eq:constrait_x1}
   {\rm{s.t.}} \  \ 0\leq x_{k,n}^{\text{p}} \leq L,\forall n\in \mathcal{N},\forall k\in \mathcal{K},
\end{equation}
\begin{equation}
\label{eq:constraint_Delta1}
  x_{k,n+1}^{\text{p}}- x_{k,n}^{\text{p}} \leq \Delta,\forall 1\leq n<N,\forall k\in \mathcal{K},
\end{equation}
\begin{equation}
\label{eq:constraint_p1}
 \sum_{k=1}^{K} \sum_{j=1}^{q_k}{p}_{j}^k\leq {P}_{\text{max}},
\end{equation}
\begin{equation}
\label{eq:constraint_qos1}
{R}^{k}_{j\rightarrow{}j}\geq R_{\text{min}}, \forall j\in D_k, \forall k\in \mathcal{K}
\end{equation}
\begin{equation}
\label{eq:constraint_sic1}
 R_{g\rightarrow{}j}^{k}\geq R^{k}_{j\rightarrow{}j}, \forall g,j\in \mathcal{D}_{k},\forall k\in \mathcal{K},\pi(g)\geq\pi(j),
\end{equation}
\end{subequations}
where $q_k$ represents the number of users assigned to the $k$-th waveguide, and $j$ is the index of the $j$-th decoded user in the cluster $\mathcal{D}_k$. To solve problem~\eqref{eq:optimization_problem_xp}, we reformulate the expression of the objective function~\eqref{eq:objective_function_initial} into a more tractable form based on WMMSE~\cite{WMMSE}. Assume that the signal $s_{j}^k$ and noise $n_{j}^k$ are independent at the user $j$, the mean square error (MSE) of user $j$ served by the $k$-th waveguide can be given by
\begin{equation}
\begin{aligned}
\label{eq:MSE}
   & e_{j}^{k}=\mathbb{E}\left[\left|v_{j}^ky_{j}^k-s_{j}^k\right|^2\right]
   \\&=\sum_{k'\neq k}\sum_{i=1}^{q_k}p_{i}^{k'}\left|v_{j}^{k}\mathbf{h}_{k',k,j}^{H}(\mathbf{x}^{\text{p}}_{k'})\mathbf{g}_{k'}(\mathbf{x}^{\text{p}}_{k'})\right|^{2}+\sigma^{2}\left|v^k_{j}\right|^{2}\\&+1-2\mathrm{Re}\left\{\sqrt{p_{j}^{k}}v_{j}^{k}\mathbf{h}_{k,k,j}^{H}(\mathbf{x}^{\text{p}}_{k})\mathbf{g}_{k}(\mathbf{x}^{\text{p}}_{k})\right\}\\&+\sum_{g=j}^{q_k}p_{g}^{k}\left|v_{j}^{k}\mathbf{h}_{k,k,j}^{H}(\mathbf{x}^{\text{p}}_{k})\mathbf{g}_{k}(\mathbf{x}^{\text{p}}_{k})\right|^{2},
\end{aligned}
\end{equation}
where $v_{j}^{k}$ represents the channel equalizer at the user $j$. The optimization problem~\eqref{eq:optimization_problem} is equivalent to the following WMMSE problem:
\begin{subequations}
\label{eq:optimization_problem2}
\begin{equation}
\label{eq:objective_function}
\min_{\mathbf{P},\mathbf{X}}\sum_{k=1}^K\sum_{j=1}^{q_k}\left(\omega_{j}^k{e}_{j}^k-\log_2\omega_{j}^k\right),
\end{equation}
\begin{equation}
\label{eq:constrait_2}
   {\rm{s.t.}} \  \eqref{eq:constrait_x1}-\eqref{eq:constraint_sic1} ,
\end{equation}
\end{subequations}
where $\omega_{j}^k$ is the weighting factor. Given other variables, the optimal solutions of  $v_{j}^k$ and $\omega_{j}^k$ are expressed as
\begin{equation}
\label{eq:best_v}
v_{j,\text{opt}}^{k}=\sqrt{p_{j}^{k}}J_{k,j}^{(-1)}\mathbf{h}_{k,k,j}^{H}(\mathbf{x}^{\text{p}}_{k})\mathbf{g}_{k}(\mathbf{x}^{\text{p}}_{k}), 
\end{equation}
\begin{equation}
\begin{aligned}
\label{eq:omega}
    \omega_{j,\text{opt}}^k=&\left(1-\frac{p_{j}^k\left|v_{j}^k\mathbf{h}_{k,k,j}^{H}(\mathbf{x}^{\text{p}}_{k})\mathbf{g}_{k}(\mathbf{x}^{\text{p}}_{k})\right|^{2}}{J_{k,j}}\right)^{-1},
\end{aligned}
\end{equation}
where $J_{k,j}=\sum_{k'\neq k}\sum_{j=1}^{q_k}p_{j}^{k'}\left|\mathbf{h}_{k',k,j}^{H}(\mathbf{x}^{\text{p}}_{k'})\mathbf{g}_{k'}(\mathbf{x}^{\text{p}}_{k'})\right|^{2}+\sum_{g=j+1}^{q_k}p_{g}^{k}\left|\mathbf{h}_{k,k,j}^{H}(\mathbf{x}^{\text{p}}_{k})\mathbf{g}_{k}(\mathbf{x}^{\text{p}}_{k})\right|^{2}+\sigma^{2}$. 

For a thorough derivation of Eq.~\eqref{eq:best_v} and Eq.~\eqref{eq:omega}, we refer the reader to~\cite[Section II-A]{WMMSE}. The above WMMSE-based reformulation converts the objective function~\eqref{eq:objective_function} into a more tractable form. However, the complex exponential terms and the reciprocal distance components in problem~\eqref{eq:MSE} introduce non-convexity in Eq.~\eqref{eq:optimization_problem2} with respect to $\mathbf{X}$ and $\mathbf{P}$. 

 To solve this problem, we introduce two sets of auxiliary variables $\theta_{k,j,i,n}$ and $u_{k,j,i,n}=\frac{\eta e^{-j\theta_{k,j,i,n}}}{\|\boldsymbol{\psi}^{\mathrm{u}}_{k,j}-\boldsymbol{\psi}^{\mathrm{p}}_{i,n}\|}$ to present the PASS-modified phase and the equivalent pinching beamforming coefficient of the $j$-th user's signal radiated by the $n$-th PA in the $i$-th waveguide, respectively. This allows the equality constraints to be expressed as the following residuals:
\begin{align}
\label{eq:b_u}
&b_{k,j,i,n}^{u}\triangleq u_{k,j,i,n}\sqrt{\left(x_{i,n}-x_{k,j}\right)^{2}+\psi_{k,j,i,n}^{2}}\notag\\&-\sqrt{\frac{\eta}{N}} e^{-j\theta_{k,j,i,n}},
\end{align}
\begin{equation}
\label{eq:b_theta}
b_{k,j,i,n}^{\theta}\triangleq\theta_{k,j,i,n}-\kappa\left(\|\boldsymbol{\psi}_{k,j}^\text{u}-\boldsymbol{\psi}_{i,n}^\text{p}\|+n_{\mathrm{eff}}x_{i,n}\right),
\end{equation}
where $\kappa=\frac{2\pi}{\lambda}$ and $\psi_{k,j,i,n}=\sqrt{\left(y_{k,j}-y_{k,n}\right)^2+d^2}$, respectively. Moreover, we define $q_{j,g}^{k,i}=\sqrt{p_{g}^{i}}\mathbf{h}_{i,k,j}^{H}(\mathbf{x}^{\text{p}}_{i})\mathbf{g}_{i}(\mathbf{x}^{\text{p}}_{i})$ as the path response of user $j$ to receive the signal of user $g$ served by waveguide $i$, which satisfies the equality constraint as follows:
\begin{equation}
\label{eq:b_q}
\begin{aligned}
b_{k,j,i,g}^{q}\triangleq q_{j,g}^{k,i}-\sqrt{p_{g}^{i}}\mathbf{u}_{k,j,i}^T\mathbf{1}_{N\times 1}, \forall j\in\mathcal{D}_k,g\in\mathcal{D}_{i},k,i\in \mathcal{K},
\end{aligned}
\end{equation}
where $\mathbf{u}^T_{k, j,i}=\left[u_{k,j,i,1},\cdots,u_{k,j,i,N}\right]\in\mathbb{C}^{1\times N}$ denotes the pinching beamforming coefficients of all PAs on the $i$-th waveguide to the user $j$. Thus, by fixing $v_{j,\text{opt}}^k$, the MSE in Eq.~\eqref{eq:objective_function} can be rewritten as a convex function with respect to ${q}_{j,g}^{k,i}$:
\begin{align}
\label{eq:obj}
&e_{j}^k=|v_{j,\mathrm{opt}}^{k}|^{2}\left(\sum_{i\neq k}\sum_{t=1}^{q_i}|q_{j,t}^{k,i}|^{2}+\sum_{g=j}^{q_k}|q_{j,g}^{k,k}|^{2}+\sigma^{2}\right)\notag\\&+1-2\mathrm{Re}\left\{\sqrt{p_{j}^k}v_{j,\mathrm{opt}}^{k}q_{j,j}^{k,k}\right\}.
\end{align}
To address the non-convexity of equality constraints~\eqref{eq:b_u},~\eqref{eq:b_theta}, and~\eqref{eq:b_q}, we exploit the PDD algorithm to alleviate the coupling constraints by utilizing the augmented Lagrangian relaxation. Specifically, by moving the equality constraints as a penalty term to~\eqref{eq:objective_function}, an augmented Lagrangian (AL) optimization problem can be constructed by
\begin{subequations}
\label{eq:optimization_problem3}
\begin{equation}
\begin{aligned}
\label{eq:objective_function_al}
&\min_{\mathbf{P},\mathbf{X},\mathbf{U},\boldsymbol{\theta},\mathbf{Q}}\sum_{k=1}^K\sum_{j=1}^{q_k}\left(\omega_{j}^k{e}_{j}^k-\log_2\omega_{j}^k\right)+\frac{1}{2\rho}\\&\times\sum_{k=1}^K\sum_{j=1}^{q_k}\left(\left\|\mathbf{B}_{k,j}^{u}+\rho\boldsymbol{\lambda}_{k,j}^{u}\right\|^{2}+\left\|\mathbf{B}_{k,j}^{\theta}+\rho\boldsymbol{\lambda}_{k,j}^{\theta}\right\|^{2}\right.\\&+\left.\left\|\mathbf{b}_{k,j}^{q}+\rho\boldsymbol{\lambda}_{k,j}^{q}\right\|^{2}\right),
\end{aligned}
\end{equation}
\begin{align}
\label{eq:constraint_qos_2}
   {\rm{s.t.}} \  {\left|q_{j,j}^{k,k}\right|^2}\geq \notag& \left(\sum_{j'=j+1}^{q_k}\left|q_{j,j'}^{k,k}\right|^2+\sum_{i\neq k}\sum_{t=1}^{q_i}\left|q_{j,t}^{k,i}\right|^2+\sigma^2\right)\\& \times r_{\text{min}}, j\in \mathcal{D}_{k},\forall k\in \mathcal{K},
\end{align}
\begin{align}
    \label{eq:constraint_sic2}
    &\frac{\left|q_{g,j}^{k,k}\right|^2}{\sum_{j'=j+1}^{q_k}\left|q_{g,j'}^{k,k}\right|^2+\sum_{i\neq k}\sum_{t=1}^{q_i}\left|q_{g,t}^{k,i}\right|^2+\sigma^2}\notag\\&\geq \frac{\left|q_{j,j}^{k,k}\right|^2}{\sum_{j'=j+1}^{q_k}\left|q_{j,j'}^{k,k}\right|^2+\sum_{i\neq k}\sum_{t=1}^{q_i}\left|q_{j,t}^{k,i}\right|^2+\sigma^2},
\end{align}
\begin{equation}
\label{eq:constrait_2}
   {\rm{s.t.}} \  \eqref{eq:constrait_x1}-\eqref{eq:constraint_p1} ,
\end{equation}
\end{subequations}
where $r_{\text{min}}=2^{R_{\text{min}}}-1$, $\mathbf{U}=\{u_{k,j,i,n}| j \in\mathcal{D}_k,k,i\in\mathcal{K},n\in \mathcal{N}\}$, $\boldsymbol{\theta}=\{\theta_{k,j,i,n}| j \in\mathcal{D}_k,k,i\in\mathcal{K},n\in \mathcal{N}\}$, $\mathbf{Q}=\{q_{j,g}^{k,i}| k,i \in\mathcal{K},j\in \mathcal{D}_k,g\in \mathcal{D}_{i}\}$, and $\rho\geq0$ is the penalty factor. Moreover, $\boldsymbol{\lambda}_{k,j}^{q}\in\mathbb{C}^{K\times N}$, $\boldsymbol{\lambda}_{k,j}^{\theta}\in\mathbb{C}^{K\times N}$, and $\boldsymbol{\lambda}_{k,j}^{q}\in\mathbb{C}^{M\times 1}$
are the Lagrangian dual variables. 

By leveraging the BCD method, we can divide the optimization
variables into four blocks, $\{\mathbf{Q},\mathbf{P}\}$, $\{\mathbf{X}\}$, $\{\mathbf{U}\}$, and $\{\boldsymbol{\theta}\}$. Each block of variables is alternatively optimized, while others remain fixed. Thus, problem~\eqref{eq:optimization_problem3} can be solved by optimizing the following four subproblems in the inner loop, while updating the Lagrangian dual variables and penalty factor in the outer loop.

\subsubsection{Subproblem with respect to $\mathbf{Q}$ and $\mathbf{P}$}
The pinching beamforming matrix $\mathbf{X}$ and the path response matrix $\mathbf{Q}$ are jointly optimized through a constrained optimization problem, while other variables are fixed. The problem \eqref{eq:optimization_problem3} can be rewritten as:
\begin{subequations}
\label{PDD_1}
\begin{equation}
    \begin{aligned}
&\min_{\mathbf{Q},\mathbf{P}}\sum_{k=1}^K\sum_{j=1}^{q_k}\Bigg{(}\omega_{j}^ke_{j}^k+\frac{1}{2\rho}\times\\&\left.\sum_{i=1}^K\sum_{g=1}^{q_i}\left(q_{j,g}^{k,i}-\sqrt{p_{g}^{i}}\mathbf{u}^T_{k,j,i}\mathbf{1}_{N\times 1}+\rho\lambda_{k,j,i,g}^{q}\right)^2\right),
    \end{aligned}
\end{equation}
\begin{equation}
\label{eq:constrait_2}
   {\rm{s.t.}} \ \eqref{eq:constraint_p1},\eqref{eq:constraint_qos_2},\eqref{eq:constraint_sic2},
\end{equation}
\end{subequations}
 Since problem~\eqref{PDD_1} is non-convex due to constraint~\eqref{eq:constraint_qos_2} and ~\eqref{eq:constraint_sic2}, we utilize the first-order Taylor expansion to approximate constraint~\eqref{eq:constraint_qos_2} as follows: 
\begin{equation}
\label{eq:sca-1}
\begin{aligned}
       \left|q_{j,j}^{k,k}\right|^2&\geq 2 \times\mathrm{Re}\left(\left(q_{j,j}^{k,k}\right)^{\left(t\right)}\left(q_{j,j}^{k,k}\right)^{*}\right)
       \\&-\left|\left(q_{j,j}^{k,k}\right)^{\left(t\right)}\right|^2\triangleq\varepsilon^{\text{min}}_{j,j},
\end{aligned}
\end{equation}
where $\left(q_{g,j}^{k,k}\right)^{\left(t\right)}$ denotes the given local point of $q_{g,j}^{k,k}$ in the $t$-th iteration of the successive convex approximation (SCA)~\cite{SCA}. 

To address the non-convex constraint~\eqref{eq:constraint_sic2}, we adopt the value of $\gamma_{j\rightarrow{}j}^k$ in the $(t-1)$-th iteration of the SCA, i.e., $\gamma_{j\rightarrow{}j}^{k,(t-1)}$, for approximating the SINR of user $j$ in the $t$-th iteration. Then, the constraint~\eqref{eq:constraint_sic2} can be transformed into a tractable constraint as follows:
\begin{equation}
    \label{eq:constraint_sic3}
    \frac{\left|q_{g,j}^{k,k}\right|^2}{\sum_{j'=j+1}^{q_k}\left|q_{g,j'}^{k,k}\right|^2+\sum_{i\neq k}\sum_{t=1}^{q_i}\left|q_{g,t}^{k,i}\right|^2+\sigma^2}\geq \gamma_{j\rightarrow{}j}^{k,(t-1)}.
\end{equation}
Similar to the constraint~\eqref{eq:constraint_qos_2}, the constraint~\eqref{eq:constraint_sic3} can be relaxed by utilizing the first-order Taylor expansion as follows:
\begin{equation}
\label{eq:sca-1}
\begin{aligned}
       \left|q_{g,j}^{k,k}\right|^2&\geq 2 \times\mathrm{Re}\left(\left(q_{g,j}^{k,k}\right)^{\left(t\right)}\left(q_{g,j}^{k,k}\right)^{*}\right)
       \\&-\left|\left(q_{g,j}^{k,k}\right)^{\left(t\right)}\right|^2\triangleq\varepsilon^{\text{min}}_{g,j}.
\end{aligned}
\end{equation}
Therefore, problem~\eqref{PDD_1} can be relaxed to the equivalent form as follows:
\begin{subequations}
\label{eq:PDD_1_2}
\begin{equation}
    \begin{aligned}
&\min_{\mathbf{Q},\mathbf{P}}\sum_{k=1}^K\sum_{j=1}^{q_k}\Bigg{(}\omega_{j}^ke_{j}^k+\frac{1}{2\rho}\times\\&\left.\sum_{i=1}^K\sum_{g=1}^{q_i}\left(q_{j,g}^{k,i}-\sqrt{p_{g}^{i}}\mathbf{u}^T_{k,g,i}\mathbf{1}_{N\times1}+\rho\lambda_{k,j,i,g}^{q}\right)^2\right),
    \end{aligned}
\end{equation}
\begin{equation}
\label{eq:constrait_qos_2_fin}
\begin{aligned}
   {\rm{s.t.}} \  &\varepsilon^{\text{min}}_{j,j}\geq r_{\text{min}}\times\left(\sum_{j'=j+1}^{q_k}\left|q_{j,j'}^{k,k}\right|^2+\sum_{i\neq k}\sum_{t=1}^{q_i}\left|q_{j,t}^{k,i}\right|^2+\sigma^2\right),\\&\forall k\in \mathcal{K},j\in \mathcal{D}_{k},
\end{aligned}
\end{equation}
\begin{equation}
\label{eq:constrait_sic_2_fin}
\begin{aligned}
     \varepsilon^{\text{min}}_{g,j}\geq& \gamma_{j\rightarrow{}j}^{k,(t-1)}\times\left(\sum_{j'=j+1}^{q_k}\left|q_{g,j'}^{k,k}\right|^2+\sum_{i\neq k}\sum_{t=1}^{q_i}\left|q_{g,t}^{k,i}\right|^2+\sigma^2\right),\\&\forall k\in \mathcal{K},j\in \mathcal{D}_{k},g\in \{j,...,q_{k}\},
\end{aligned}
\end{equation}
\begin{equation}
\label{eq:constrait_2}
    \eqref{eq:constraint_p1}.
\end{equation}
\end{subequations}
Problem~\eqref{eq:PDD_1_2} is a convex optimization problem and it can be solved via CVX~\cite{cvx}.

\subsubsection{Subproblem with respect to $\mathbf{X}$}
The subproblem for optimizing the pinching beamforming matrix $\mathbf{X}$ is given by
\begin{subequations}
\begin{equation}
\begin{aligned}
&\min_{\mathbf{X}}\sum_{k=1}^K\sum_{j=1}^{q_k}\frac{1}{2\rho}\left(\left\|\mathbf{B}_{k,j}^{u}+\rho\boldsymbol{\lambda}_{k,j}^{u}\right\|^{2}+\left\|\mathbf{B}_{k,j}^{\theta}+\rho\boldsymbol{\lambda}_{k,j}^{\theta}\right\|^{2}\right),
\end{aligned}\end{equation}
\begin{equation}
\label{eq:constrait_X_else}
   {\rm{s.t.}} \  \ ~\eqref{eq:constrait_x},~\eqref{eq:constraint_Delta}.
\end{equation} 
\end{subequations}
\setcounter{TempEqCnt}{\value{equation}} 
\setcounter{equation}{21} 
\begin{figure*}[ht]
    \begin{equation}
    \begin{aligned}
          \label{eq:func_al}
          &L^{\text{AL}}_{k,j,i,n}(x_{i,n}) = \frac{1}{2\rho} \left|u_{k,j,i,n}^k\sqrt{(x_{i,n}-x_{k,j}^\mathrm{u})^2+\psi_{k,j,i,n}^2} \right.\left.-\sqrt{\frac{\eta}{N}} e^{-j\theta_{k,j,i,n}}+\rho\lambda^{u}_{k,j,i,n} \right|^2\\&+\frac{1}{2\rho}\left|\theta_{k,j,i,n}-\kappa\left(\sqrt{(x_{i,n}-x_{k,j}^\mathrm{u})^2+\psi_{k,j,i,n}^2}-n_{\text{eff}}x_{i,n}\right)+\rho\lambda^{\theta}_{k,j,i,n}\right|^2,     
    \end{aligned}
    \end{equation}
        \begin{equation}
    \begin{aligned}
          \label{eq:func_al2}
          &L^{\text{AL}}_{k,j,i,n}(x_{i,n}) = \underbrace{\frac{1}{2\rho}\left(u_{k,j,i,n}^2+\kappa\right)\left(x_{i,n}-x_{k,j}^\mathrm{u}\right)^2+\frac{1}{2\rho}\kappa^2n_{\text{eff}}x_{i,n}^2-\kappa\left(\lambda_{k,j,i,n}^{\theta}+\frac{\theta_{k,j,i,n}}{\rho}\right)n_{\text{eff}}x_{i,n}}_{\hat{L}^{\text{AL}}_{k,j,i,n}(x_{i,n})}\\&+\underbrace{\xi_{k,j,i,n}\sqrt{(x_{i,n}-x_{k,j}^\mathrm{u})^2+\psi_{k,j,i,n}^2}+\frac{\kappa^2n_{\text{eff}}}{\rho}x_{i,n}\sqrt{(x_{i,n}-x_{k,j}^\mathrm{u})^2+\psi_{k,j,i,n}^2}}_{L^\text{NC}_{k,j,i,n}(x_{i,n})},       
    \end{aligned}
    \end{equation}
    \hrulefill
\end{figure*}
\addtocounter{equation}{0}
 \begin{flushleft}The AL term $\left\|\mathbf{B}_{k,j}^{u}+\rho\boldsymbol{\lambda}_{k,j}^{u}\right\|^{2}+\left\|\mathbf{B}_{k,j}^{\theta}+\rho\boldsymbol{\lambda}_{k,j}^{\theta}\right\|^{2}$  can be
 \end{flushleft}
 \noindent
reformulated as given in Eq.~\eqref{eq:func_al}, shown at the top of the next page. Subsequently, by extracting the components involving the variable
$\mathbf{X}$, Eq.~\eqref{eq:func_al} can be further rewritten as Eq.~\eqref{eq:func_al2}, where $\xi_{k,j,i,n}=\mathrm{Re}\left\{u_{k,j,i,n}^H\left(\lambda_{k,j,i,n}^u-\frac{1}{\rho}\sqrt{\frac{\eta}{N}} e^{-i\theta_{k,j,i,n}}\right)\right.-\left.\kappa\left(\lambda_{k,j,i,n}^{\theta}
+\frac{\theta_{k,j,i,n}}{\rho}\right)\right\}$. Note that Eq.~\eqref{eq:func_al2} consists of a convex component $\hat{L}^{\text{AL}}_{k,j,i,n}(x_{i,n})$ and a nonconvex component $L^{\text{NC}}_{k,j,i,n}(x_{i,n})$. Specifically, the nonconvex function $L^{\text{NC}}_{k,j,i,n}$ in Eq.~\eqref{eq:func_al2} consisted of two functions $f^{(1)}_{k,j,i,n}(x_{i,n})=\frac{\kappa^2n_{\text{eff}}}{\rho}x_{i,n}\sqrt{(x_{i,n}-x_{k,j}^\mathrm{u})^2+\psi_{k,j,i,n}^2}$ and $f^{(2)}_{k,j,i,n}(x_{i,n})=\xi_{k,j,i,n}\sqrt{(x_{i,n}-x_{k,j}^\mathrm{u})^2+\psi_{k,j,i,n}^2}$. For $f^{(1)}_{k,j,i,n}(x_{i,n})$, we utilize Jensen's inequality to obtain an upper bound as follows:

\begin{equation}
\begin{aligned}
    \label{eq:Jensen_equality}
    &f^{(1)}_{k,j,i,n}(x_{i,n})\leq \\& \frac{x_{i,n}\left[\left(x_{i,n}-x_{k,j}^\mathrm{u}\right)^2+\left(\left(x_{i,n}^{(t)}-x_{k,j}^\mathrm{u}\right)^2+2\psi^2_{k,j,i,n}\right)\right]}{2\sqrt{\left(x_{i,n}^{(t)}-x_{k,j}^\mathrm{u}\right)^2+\psi^2_{k,j,i,n}}}.
\end{aligned}
\end{equation}
Since the RHS of Eq.~\eqref{eq:Jensen_equality} is still nonconvex, we employ the SCA method to make the convex approximation. Specifically, for the nonconvex component $x_{i,n}\left(x_{i,n}-x_{k,j}^\mathrm{u}\right)^2=x_{i,n}^3-2x_{k,j}^\mathrm{u}x_{i,n}^2+\left(x_{k,j}^\mathrm{u}\right)^2x_{i,n}$, by exploiting first-order Taylor expansion, we can get the upper bound of the term as follows:
\begin{equation}
    \label{eq:f1_first_Taylor}
    -2x_{k,j}^\mathrm{u}x_{i,n}^2\leq-2x_{k,j}^\mathrm{u}\left(x_{i,n}^{\left(t-1\right)}\right)^2-4x_{k,j}^\mathrm{u}\left(x_{i,n}^{\left(t-1\right)}\right)^2x_{i,n}.
\end{equation}
Then, the Eq.~\eqref{eq:Jensen_equality} can be rewritten as:
\begin{align}
    \label{eq:Jensen_equality_fin}
    f^{(1)}_{k,j,i,n}(x_{i,n})&\leq \frac{x_{i,n}^3+s_{k,j,i,n}x_{i,n}+2x_{k,j}^\mathrm{u}\left(x_{i,n}^{\left(t-1\right)}\right)^2}{2\sqrt{\left(x_{i,n}^{(t)}-x_{k,j}^\mathrm{u}\right)^2+\psi^2_{k,j,i,n}}}\notag\\&=\hat{f}^1_{k,j,i,n}(x_{i,n})  
\end{align}
where $s_{k,j,i,n}=\left(x_{k,j}^\mathrm{u}\right)^2+2\psi_{k,j,i,n}^2+\left(x_{i,n}^{\left(t-1\right)}-x_{k,j}^\mathrm{u}\right)^2-4x_{k,j}^\mathrm{u}x_{i,n}^{\left(t-1\right)}$. The complexity for solving optimization problem~\eqref{eq:optimization_problem2} is governed by the component $f^{(2)}_{k,j,i,n}(x_{i,n})$. Again, by utilizing the first-order Taylor expansion of $f^{(2)}_{k,j,i,n}(x_{i,n})$ as its tight upper bound, $f^{(2)}_{k,j,i,n}(x_{i,n})$ can be rewritten as:
\begin{equation}
\begin{aligned}
\label{eq:L_DC2}
  & f^{(2)}_{k,j,i,n}(x_{i,n}) \leq \frac{1}{2\rho}\left(1+u_{k,j,i,n}^2\right)\left(x_{i,n}-x_{k,j}^\mathrm{u}\right)^2\\&+\frac{\xi_{k,j,i,n}\left(x_{i,n}^{\left(t-1\right)}-x_{k,j}^\mathrm{u}\right)}{\left(x_{i,n}^{\left(t-1\right)}-x_{k,j}^\mathrm{u}\right)^2+\psi_{k,j,i,n}^2}\left(x_{i,n}-x_{i,n}^{\left(t-1\right)}\right)\\&=\hat{f}^{\left(2\right)}_{k,j,i,n}(x_{i,n}),   
\end{aligned}
\end{equation}
where $x_{i,n}^{\left(t\right)}$ is the solution of $x_{i,n}$ obtained at the $t$-th iteration. Finally, $\mathbf{X}$ can be updated by solving the following problem:
\begin{subequations}
\label{eq:PDD-x-fin}
\begin{equation}
\begin{aligned}
&\min_{\mathbf{X}}\sum_{k\in\mathcal{K}}\sum_{j\in\mathcal{D}_k}\sum_{i\in\mathcal{K}}\sum_{n\in \mathcal{N}}\widehat{L}_{k,j,i,n}^{\mathrm{AL}}\left(x_{i,n}\right)\\&+\sum_{k\in\mathcal{K}}\sum_{j\in\mathcal{D}_k}\sum_{i\in\mathcal{K}}\sum_{n\in \mathcal{N}}\widehat{L}_{k,j,i,n}^{\mathrm{NC}}\left(x_{i,n}\right)
\end{aligned}\end{equation}
\begin{equation}
\label{eq:constrait_X_else}
   {\rm{s.t.}} \  \ ~\eqref{eq:constrait_x},~\eqref{eq:constraint_Delta}.
\end{equation}
\end{subequations}
where $\widehat{L}_{j,k,n}^{\mathrm{NC}}\left(x_{i,n}\right)=\hat{f}^{(1)}_{k,j,i,n}(x_{i,n})+\hat{f}^{(2)}_{k,j,i,n}(x_{i,n})$.
Then, the problem~\eqref{eq:PDD-x-fin} can be solved by the interior point method, which utilizes the CVX tool. 

\subsubsection{Subproblem with respect to $\mathbf{U}$}
The subproblem for solving $\mathbf{U}$ is given by
\begin{equation}
    \begin{aligned}
    \label{eq:u}
&\min_{\mathrm{U}}\sum_{k=1}^K\sum_{j=1}^{q_k}\sum_{i=1}^K\bigg{(}\left\|\mathbf{R}_{k,j,i}\mathbf{u}_{k,j,i}+\boldsymbol{\zeta}_{k,j,i}\right\|^2\\&\left.+\sum_{g=1}^{q_k}\left\|{q}^{k,i}_{j,g}+\rho\lambda_{k,j,i,g}^{q}-\sqrt{p_{g}^{i}}{(\mathbf{u}_{k,j,i})}^T\mathbf{1}_{N\times 1}\right\|^2\right),      
\end{aligned}
\end{equation}
where $\mathbf{R}_{k,j,i}=\mathrm{diag}(r_{k,j,i,1},\ldots,r_{k,j,i,N})\in\mathbb{R}^{N\times N}$, $\boldsymbol{\zeta}_{k,j,i}=\left[\zeta_{k,j,i,1},...,\zeta_{k,j,i,N}\right]^T\in\mathbb{C}^{N\times 1}$, $r_{k,j,i,n}=\|\boldsymbol{\psi}^{\mathrm{u}}_{k,j}-\boldsymbol{\psi}^{\mathrm{p}}_{i,n}\|$, and $\zeta_{k,j,i,n}=\rho\lambda^u_{k,j,i,n}-\sqrt{\frac{\eta}{N}} e^{-\theta_{k,j,i,n}}$. Hence, the closed-form solution of $\mathbf{u}_{k,j,i}$, is derived as
\begin{equation}
\begin{aligned}
&\mathbf{u}_{k,j,i}=\left(\mathbf{R}_{k,j,i}^{H}\mathbf{R}_{k,j,i}+\left(\sum_{g=1}^{q_k}p_{g}^{i}\right)\mathbf{1}_{N\times N}\right)^{-1}\\&\times\left(\sum_{g=1}^{q_k}\sqrt{p_{g}^{i}}\left(q_{j,g}^{k,i}+\rho\lambda_{k,j,i,g}^{q}\right)\mathbf{1}_{N\times 1}\right.-\mathbf{R}_{k,j,i}^{H}\boldsymbol{\zeta}_{k,j,i}\Bigg{)}.   
\end{aligned}
\end{equation}

\subsubsection{Subproblem with respect to $\boldsymbol{\theta}$}
The signal phases $\boldsymbol{\theta}$ can be updated by solving the unconstrained convex optimization problem:
\begin{equation}
\label{eq:PDD-4}
\begin{aligned}
&\min_{\theta} \sum_{k=1}^K\sum_{j=1}^{q_k} \sum_{i=1}^K \sum_{n=1}^N \frac{1}{2\rho} \left[ \theta_{k,j,i,n} - \kappa \left(r_{k,j,i,n} + n_{\text{eff}} x_{i,n}\right)\right.\\&\left.+ \rho\lambda^{\theta}_{k,j,i,n} \right]^2+L^{\text{EC}}_{k,j,i,n}\left(\theta_{k,j,i,n}\right),
\end{aligned}
\end{equation}
where $L^{\text{EC}}_{k,j,i,n}\left(\theta_{k,j,i,n}\right)$ is given by
\begin{equation}
\begin{aligned}
    \label{eq:L_exp}
    &L^{\text{EC}}_{k,j,i,n}\left(\theta_{k,j,i,n}\right)=\\&-\sqrt{\frac{\eta}{N}}\text{Re}\left\{e^{j\theta_{k,j,i,n}}\left(\lambda^u_{k,j,i,n}+\frac{1}{\rho}u_{k,j,k,m}r_{k,j,i,n}\right)\right\}.
\end{aligned}
\end{equation}
Due to the complex component $L^{\text{EC}}_{k,j,i,n}\left(\theta_{k,j,i,n}\right)$ in Eq.~\eqref{eq:PDD-4}, we construct Lipschitz gradient surrogate of $L^{\text{EC}}_{k,j,i,n}\left(\theta_{k,j,i,n}\right)$, which is defined as follows.
\begin{definition}
If the gradient of function $f\left(\mathbf{x}\right)$ is $\varrho$-Lipschitz continuous over a set $\mathbf{X}$, we have 
 \begin{equation}
 \label{eq:Lipschitz gradient constant}
     \|\nabla f\left(\mathbf{x}_2\right)-\nabla f\left(\mathbf{x}_1\right)\|\leq\varrho\|\mathbf{x}_2-\mathbf{x}_1\|,\forall \mathbf{x}_1,\mathbf{x}_2\in\mathbf{X},
 \end{equation}
 where $\varrho$ is the Lipschitz gradient constant~\cite{surrogate}. Then, the Lipschitz gradient surrogate of the function $f\left(\mathbf{x}_1\right)$ can be given by
  \begin{equation}
  \begin{aligned}
 \label{eq:Lipschitz gradient surrogate}
     &g\left(\mathbf{x_1},\mathbf{x}_1^{\left(t-1\right)}\right)=f\left(\mathbf{x}_1^{\left(t-1\right)}\right)
     \\&+\nabla_{\mathbf{x}_1}f\left(\mathbf{x}_1^{\left(t-1\right)}\right)^T\left(\mathbf{x}_1-\mathbf{x}_1^{\left(t-1\right)}\right)+\frac{\varrho}{2}\left\|\left(\mathbf{x}_1-\mathbf{x}_1^{\left(t-1\right)}\right)\right\|^2.
  \end{aligned}
 \end{equation}
\end{definition}

 Since the gradient of $L^{\text{EC}}_{k,j,i,n}\left( \theta \right)$ is $\varrho$-Lipschitz continuous, $L^{\text{EC}}_{k,j,i,n}\left( \theta \right)$ is upper bound by utilizing the Lipschitz gradient surrogate as follows:
\begin{equation}
\begin{aligned}
\label{eq:lgs}
 &{L}_{k,j,i,n}^{\text{EC}} \left( \theta \right)\leq\hat{L}_{k,j,i,n}^{\text{EC}} \left( \theta \right)= L_{k,j,i,n}^{\text{EC}} \left( \theta_{k,j,i,n}^{(t-1)} \right) 
 \\&+\nabla_ \theta \cdot {L}_{k,j,i,n}^{\text{EC}} \left( \theta_{k,j,i,n}^{(t-1)} \right) \left( \theta_{k,j,i,n} - \theta_{k,j,i,n}^{(t-1)} \right)
 \\&+ \frac{\varrho_{k,j,i,n}}{2} \left( \theta_{k,j,i,n} - \theta_{k,j,i,n}^{(t-1)} \right)^2,       
\end{aligned}
\end{equation}
where $\varrho_{k,j,i,n}$ is the Lipschitz gradient constant and $\theta_{k,j,i,n}^{(t)}$ denotes the optimital solution of $\theta_{k,j,i,n}$ in the $t$-th iteration. Specifically, $\varrho_{k,j,i,n}$ can be given by 
\begin{equation}
    \label{eq:varrho}
    \varrho_{k,j,i,n}=\sqrt{\frac{\eta}{N}}\left|\lambda^u_{k,j,i,n}+\frac{1}{\rho}u_{k,j,i,n}r_{k,j,i,n}\right|,
\end{equation}
and the derivative of the function ${L}_{k,j,i,n}^{\text{EC}} \left( \theta_{k,j,i,n} \right)$ is given by
\begin{equation}
\begin{aligned}
    \label{eq：L_exp}
    &\nabla_ \theta \cdot {L}_{k,j,i,n}^{\text{EC}} \left( \theta \right)=\sqrt{\frac{\eta}{N}} \text{Re}\left\{\lambda_{k,j,i,n}^u+\frac{1}{\rho}u_{k,j,i,n}r_{k,j,i,n}\right\}\\&\times\sin\theta+\sqrt{\frac{\eta}{N}} \text{Im}\left\{\lambda_{k,j,i,n}^u+\frac{1}{\rho}u_{k,j,i,n}r_{k,j,i,n}\right\}\cos\theta.    
\end{aligned}
\end{equation}
For brevity, the derivation of $\varrho_{k,j,i,n}$is omitted here. The reader can be referred to~\cite{rho} for more details. Then, the optimal solution of $\theta_{k,j,i,n}$ is given by:
\begin{equation}
\begin{aligned}
\label{eq:theta_opt}
&\theta_{k,j,i,n}=\frac{1}{\varrho_{k,j,i,n}+\frac{1}{\rho}}\times\left(\varrho_{k,j,i,n}\theta_{k,j,i,n}^{(t-1)}-\lambda_{k,j,i,n}^{\theta}\right.\\&\left.+\frac{\kappa}{\rho}(r_{k,j,i,n}+n_{\mathrm{eff}}x_{i,n})\varrho_{k,j,i,n}-\nabla_{\theta}L^{\text{EC}}\left(\theta_{k,j,i,n}^{(t-1)}\right)\right).
\end{aligned}
\end{equation}
\begin{algorithm}[!tp]
\caption{PDD Algorithm for Power Allocation and Pinching Beamforming}
\label{alg:pdd}
\textbf{Input:} User scheduling index $\mathcal{C}$.
\begin{algorithmic}[1]
\STATE Initialize variable matrix $\mathbf{X}^{\left(0\right)}$, $\mathbf{P}^{\left(0\right)}$, $\mathbf{Q}^{\left(0\right)}$, and $\mathbf{U}^{\left(0\right)}$. Initialize dual variables $\boldsymbol{\lambda}^{\left(0\right)} = \mathbf{0}$ and penalty parameter $\rho^{\left(0\right)}$.
\STATE  Set iteration index $t = 0$.
\REPEAT
    \WHILE{the fractional decrement of the AL function~\eqref{eq:objective_function_al} exceeds the threshold $\epsilon_{\text{1}}$}
        \STATE Update $v^k_{j}$ and $\omega^k_{j}$ with~\eqref{eq:best_v} and~\eqref{eq:omega}.
        \STATE Update $\{\mathbf{P}^{\left(t+1\right)} , \mathbf{Q}^{\left(t+1\right)} \}$ by solving problem~\eqref{eq:PDD_1_2}.
        \STATE Update $\mathbf{X}^{\left(t+1\right)} $ by solving problem~\eqref{eq:PDD-x-fin}.
        \STATE Update $\mathbf{U}^{\left(t+1\right)} $ with~\eqref{eq:u}.
        \STATE Update $\boldsymbol{\theta}^{\left(t+1\right)} $ by solving problem~\eqref{eq:theta_opt}.
    \ENDWHILE
    \IF{$\| \mathbf{B}^{(t)} \|_{\infty} \le 0.8 \| \mathbf{B}^{(t-1)} \|_{\infty}$}
        \STATE Update Lagrangian dual variables $\boldsymbol{\lambda}^{(t+1)} = \frac{\mathbf{B}^{(t)}}{\rho^{(t)}} +\boldsymbol{\lambda}^{(t)}  $.
    \ELSE
        \STATE Update penalty factor $\rho^{(t+1)} = 0.8\times\rho^{(t)} $.
    \ENDIF
    \STATE $t=t+1$.
    \UNTIL The maximum residual $\| \mathbf{B}^{(t)} \|_{\infty}$ is below the threshold $\epsilon_{\text{2}}$.
\end{algorithmic}
\textbf{Output:} $\mathbf{P}$ and $\mathbf{X}$.
\end{algorithm}

Following the above discussions, the proposed PDD algorithm is exploited to address the original problem~\eqref{eq:optimization_problem_xp}. Specifically, the AL problem~\eqref{eq:optimization_problem3} is optimized by alternatively solving the aforementioned four subproblems with a BCD manner in the inner loop, where the SCA and the Lipschitz gradient surrogate methods are utilized to handle non-convexity in the problem. The Lagrangian dual variables and the penalty factors are updated to enforce constraint satisfaction in the outer loop. This iterative process ensures that the AL function is monotonically non-increasing in each iteration, which guarantees the overall algorithm converges to a stationary point. The details of the PDD algorithm are shown in \textbf{Algorithm~\ref{alg:pdd}}.

\subsection{Matching Algorithm for User Scheduling}
In this subsection, we propose a low-complexity matching algorithm for user scheduling. This algorithm consists of two steps: initialize the user-to-waveguide assignment, and update the user scheduling scheme in a loop through the matching algorithm. 

We first propose an initialization algorithm to generate a feasible initial user-to-waveguide assignment for the subsequent matching process. Specifically, initialization first establishes preference list for user set $\mathcal{M}$ and waveguide set $\mathcal{K}$ based on channel gain. Subsequently, an iterative matching process is employed, where each unassigned user is assigned to its most preferred waveguide. This procedure is repeated until each waveguide accepts the best-ranking users up to its designated quota, thereby forming a complete initial user scheduling scheme for subsequent optimization. The detail of the initialization algorithm is given in \textbf{Algorithm~\ref{alg:initial}}.
\begin{algorithm}[!tp]
    \renewcommand{\algorithmicrequire}{\textbf{Input:}}
    \renewcommand{\algorithmicensure}{\textbf{Output:}}
  \caption{Initialization Algorithm }
   \textbf{Input:} The power allocation coefficients matrix $\mathbf{P}$ and the PAs positions matrix $\mathbf{X}$.
  \begin{algorithmic}[1]
    \label{alg:initial}
       \STATE  Initialize the preference lists between the user sets and the waveguide sets based on the channel gain $|\mathbf{h}_{k,m}^T\left(\mathbf{x}^{\text{p}}_{k}\right)\mathbf{g}\left(\mathbf{x}^{\text{p}}_{k}\right)|^2,m\in\mathcal{M},k\in\mathcal{K}$.
        \STATE Initial the set of accepted users by waveguide $k$ $\mathcal{A}^0\left(k\right)=\emptyset$, the rejected users $\mathcal{R}^0\left(k\right)=\emptyset$ and the set of rejected waveguides $\mathcal{R}^0\left(m\right)=\emptyset$.
        \STATE Set iteration index $j = 0$;
      \REPEAT       
              \STATE $j =j+1$;
              \STATE Assign all users who have not yet been assigned  $m\in\mathcal{M}\backslash \cup_{k\in\mathcal{K}}\mathcal{A}^{j-1}$ to their current best user set that has not rejected them.
              \STATE Denote the users who are assigned to waveguide $k$ as $\hat{m}_1^{k},...,\hat{m}_{s'}^{k}$.
              \STATE Waveguide $k$ accepts the first $q_k$ best ranked users from $\mathcal{S}=\{\hat{m}_1^{k},...,\hat{m}_{s}^{k}\}=\mathcal{A}^{j-1}\left(k\right)\cup\{\hat{m}_1^{k},...,\hat{m}_{s'}^{k}\}$, where $s$ denotes the number of candidate users proposing to the waveguide $k$.
              \STATE Update the rejected user set $\mathcal{R}^{j}\left(k\right)=\{\hat{m}_{q_k+1}^{k},...,\hat{m}_{s}^{k}\}$ and the set of rejected waveguides $\mathcal{R}^{j}\left(m\right)=\{k\in\mathcal{K}:m\in\mathcal{R}^{j}(k)\}$.
      \UNTIL All waveguides achieve their maximum number of users.
  \end{algorithmic}
  \textbf{Output:} $\Phi$.
\end{algorithm}

Then, we formulate the user scheduling problem as a many-to-one matching problem, which aims to maximize the all users' sum rate. By fixing the power allocation coefficients $\mathbf{P}$ and the PA positions $\mathbf{X}$, the original problem~\eqref{eq:optimization_problem} is transformed into
\begin{subequations}
\label{eq:optimization_problem_cpi}
\begin{equation}
\label{eq:objective_function_match}
    \max_{\mathcal{C}}\sum_{k=1}^K\sum_{m=1}^M{R}^{k}_{m\rightarrow{}m},
\end{equation}

\begin{equation}
\label{eq:constraint_num2}
{\rm{s.t.}} \  \sum_{m=1}^Mc_m^k\leq \bar{q},\forall k\in \mathcal{K},
\end{equation}
\begin{equation}
\label{eq:constraint_c2}
\sum_{k=1}^Kc_m^k=1, \forall m\in \mathcal{M}.
\end{equation}
\end{subequations}
Constraints~\eqref{eq:constraint_num2} and~\eqref{eq:constraint_c2} restrict that, each waveguide can serve multiple users simultaneously, while each user is assigned to a single waveguide. Thus, the relationship between the users and waveguides can be regarded as a many-to-one matching, which is defined as follows.

\begin{definition}
 A many-to-one matching $\Phi$ is a function that maps from the  user set $\mathcal{M}$ to the waveguide set $K$, satisfying the following conditions:
\begin{enumerate}
    \item $\Phi(m)\in \mathcal{K},\forall m \in \mathcal{M}$;
    \item $|\Phi(k)|= q_k,\forall k \in \mathcal{K}$;
    \item $|\Phi(m)|\leq1, \forall m\in \mathcal{M}$;
    \item $m\in \Phi(k)\Leftrightarrow \Phi(m)=k$.
\end{enumerate}
\end{definition}
\noindent
The conditions in Definition~2 indicate the matching relationship between the users and the waveguides. Note that if the parameter is the user index $m$, $\Phi(m)$ denotes the matched waveguide index of the user $m$. When the parameter is a waveguide index $k$, $\Phi(k)$ represents the user cluster served by the $k$-th waveguide.

We introduce the preference value for the user $m$ served by waveguide $k$, which denotes the achievable data rate of the user $m$ served by waveguide $k$. This preference value is defined as
\begin{equation}
\mathcal{H}_m\left(\Phi\right)=\log_2{(1+\gamma_{m\rightarrow{}m}^k)}.
\end{equation}
Similarly, the preference value of the waveguide $k$ can be defined as the sum rate of the user cluster it serves, which can be expressed as:
\begin{equation}
\mathcal{H}_{k}\left(\Phi\right)=\sum_{m\in \mathcal{D}_k}\mathcal{H}_m\left(\Phi\right).
\end{equation}
Based on the preference value, every user has a preference ordering $\succ_{m}$ over the waveguide set $\mathcal{K}$. Specifically, for a given user $m$, any two waveguides $k$ and $k'$, the preference ordering is defined as
\begin{equation}
\label{eq:change_user}
\left(k,\Phi\right)\succ_{m}\left(k',\Phi'\right)\Leftrightarrow\mathcal{H}_{m}\left(\Phi\right)>\mathcal{H}_{m}\left(\Phi'\right),
\end{equation}
which implies that user $m$ prefers waveguides $k$ in matching $\Phi$ to waveguide $k'$ in matching $\Phi'$ when the achievable rate for user $m$ on waveguide $k$ is greater than that on waveguide $k'$. Similarly, each waveguide has a preference ordering $\succ_{k}$, which denotes the preference relation between the waveguide and the user set. For any two subsets of users $\mathcal{D}_k$ and $\mathcal{D}_{k'}$ and any two matchings $\Phi$ and $\Phi'$, the preference ordering is introduced as follows:
\begin{equation}
\label{eq:change_waveguide}
\left(\mathcal{D}_k,\Phi\right)\succ_{k}\left(\mathcal{D}_{k'},\Phi'\right)\Leftrightarrow\mathcal{H}_{k}\left(\Phi\right)>\mathcal{H}_{{k'}}\left(\Phi'\right),
\end{equation}
which means that waveguide $k$ prefers user set $\mathcal{D}_{k}$ due to its higher achievable sum rate. 

It can be found that the preference values and preference orderings of these two distinct participant sets are highly coupled~\cite{externality,externality2}. This phenomenon arises from externalities, where a user's achievable rate is determined not only by the interference from members within the same cluster but also by the interference generated by users in other clusters. Consequently, the many-to-one matching problem is denoted as a two-sided matching problem with externalities.

For solving this problem, we propose a low-complexity matching algorithm to obtain a two-sided exchange stable matching. To define exchange stability, we first introduce a swap matching $\Phi_m^j$ where user $m$ and user $j$ exchange their assigned waveguides, while other user-to-waveguide assignments remain unchanged. The swap matching among users is defined as follows. 
\begin{algorithm}[!tp]
\caption{Matching Algorithm for User Scheduling}
\label{alg:user_scheduling}
\textbf{Input:} The initial matching $\Phi$, which is obtained by~\textbf{Algorithm~\ref{alg:initial}}.
\begin{algorithmic}[1]
    \REPEAT
    \FOR{any user $m \in M$}
        \STATE Select a candidate user $j$.
        \IF{user $(m, j)$ constitutes a swap-blocking pair}
            \STATE  Update matching $\Phi = \Phi_m^j$.
        \ELSE
            \STATE The matching scheme $\Phi$ remains unchanged.
        \ENDIF
    \ENDFOR
    \UNTIL{no swap-pair exists.}
\end{algorithmic}
\textbf{Output:} The stable matching $\Phi$.
\end{algorithm}
\begin{definition}
    For any two users $m$ and $j$, the swap matching is given by
    \begin{equation}
        \Phi_m^j=\left\{\Phi\backslash\left\{\left(m,k\right),\left(j,i\right)\right\}\cup\left\{\left(j,k\right),\left(m,i\right)\right\}\right\},\notag
    \end{equation}
    where $\Phi\left(m\right)=k$ and $\Phi\left(j\right)=i$.
\end{definition}
\noindent
Specifically, each swap matching operation is restricted to a pair of users and two associated waveguides. Accordingly, the swap-blocking pair is introduced to approve a swap operation, which is defined as follows.
\begin{definition}
\label{def:swapblocking}
     Given a pair of users $\left(m,j\right)$ and a matching function $\Phi$, where user $m$ is served by waveguide $k$ $\left(\Phi\left(m\right)=k\right)$ and user $j$ is served by waveguide $i$ $\left(\Phi\left(j\right)=i\right)$, the user pair $\left(m,j\right)$ is denoted as a swap-blocking pair when the following conditions are satisfied.
\begin{enumerate}
    \item The performances of all involved users and the user sets are non-decreasing, i.e.,
    \begin{equation}
        \forall x\in\{m,j,k,i\}, \mathcal{H}_{x}\left(\Phi_m^j\right)\geq \mathcal{H}_{x}\left(\Phi\right).\notag
    \end{equation}
    \item At least one involved user or user set obtains a performance improvement, i.e.,
        \begin{equation}
        \exists x\in\{m,j,k,i\}, \mathcal{H}_{x}\left(\Phi_m^j\right)> \mathcal{H}_{x}\left(\Phi\right).\notag
    \end{equation}
\end{enumerate}
\end{definition}
From Definition~\ref{def:swapblocking}, the swap-blocking pair properties guarantee that any approved swap matching will either maintain or improve the achievable rates for all involved users, with at least one user obtaining a rate enhancement. Based on the above definitions, we propose a matching algorithm to optimize user scheduling, as shown in the \textbf{Algorithm~\ref{alg:user_scheduling}}. Specifically, every two users served by different waveguides are formed as a candidate swap-blocking pair. The BS then verifies whether exchanging their matches can improve the rate of at least one participant without degrading the individual rates of the involved users or the sum rates of their respective clusters. By carrying out a sequence of valid swap matching operations, the matching will eventually achieve a novel  two-sided exchange-stable (TES) stability, which can be defined as follows. 
\begin{definition}
\label{def:TES}
 A matching $\Phi$ is TES when no swap blocking pair can be found within it.
\end{definition}
\noindent
The TES matching implies that no swap-blocking pairs exist, which means that no user and waveguide can exchange assignments to increase their performance~\cite{TES,matching}. 

\begin{algorithm}[!tp]
\caption{Matching-PDD Algorithm}
\label{alg:matching-pdd}
\textbf{Input:} Users' locations $\boldsymbol{\psi}^{\text{u}}$ and feed points' locations $\boldsymbol{\psi}^{\text{feed}}$.
\begin{algorithmic}[1]
 \STATE   Initial power allocation coefficient matrix $\mathbf{P}$ and pinching deployment matrix $\mathbf{X}$.
\STATE Get the initial matching $\Phi$ via \textbf{Algorithm~\ref{alg:initial}}.
\STATE Update the matching $\Phi$ via \textbf{Algorithm~\ref{alg:user_scheduling}}.
\STATE Update $\mathbf{P}$ and  $\mathbf{X}$ via \textbf{Algorithm~\ref{alg:pdd}}.
\end{algorithmic}
  \textbf{Output:} $\mathbf{P}$, $\mathbf{X}$, and $\mathcal{C}$.
\end{algorithm}
\subsection{Overall Algorithm}
For solving the original problem~\eqref{eq:optimization_problem}, we propose a Matching-PDD algorithm. Specifically, the appropriate user-to-waveguide assignment is obtained through the matching algorithm in \textbf{Algorithm~\ref{alg:user_scheduling}}. Then, the power allocation coefficients $\mathbf{P}$ and PAs positions $\mathbf{X}$ are iteratively optimized through the PDD algorithm in \textbf{Algorithm~\ref{alg:pdd}}. We summarize the overall algorithm for solving the problem~\eqref{eq:optimization_problem}, which is given in \textbf{Algorithm~4}.

 Then, we analyze the computational complexity of \textbf{Algorithm 4}. For solving problem~\eqref{eq:optimization_problem},  the computational complexity is dominated by utilizing CVX tool to optimize the convex subproblems. For solving subproblem~\eqref{eq:PDD_1_2}, the computational complexity is given by $\mathcal{O}(I_{\text{iter,1}}M^{3.5})$, where $I_{\text{iter,1}}$ is the number of iterations in the SCA method for convergence. Similarly, the complexity for solving subproblem~\eqref{eq:PDD-x-fin},~\eqref{eq:u} and ~\eqref{eq:PDD-4} are given by $\mathcal{O}\left(I_{\text{iter,2}}\left(KN\right)^{1.5}\right)$, $\mathcal{O}\left(MKN^{3}\right)$, and $\mathcal{O}\left(I_{\text{iter,4}}MKN\right)$, where $I_{\text{iter,2}}$ and $I_{\text{iter,4}}$ denote the number of iterations in subproblem~\eqref{eq:PDD-x-fin} and ~\eqref{eq:PDD-4}. Thus, the complexity of \textbf{Algorithm~1} is expressed as $O\left(I_{\mathrm{in}}I_{\mathrm{out}}\left(I_{\text{iter,1}}M^{3.5}+I_{\text{iter,2}}\left(KN\right)^{1.5}+MKN^{3}\right.\right.$ $\left.\left.+I_{\text{iter,4}}MKN\right)\right)$, where  $I_{\mathrm{in}}$ and $I_{\mathrm{out}}$ denote the number of iterations in the inner loop and outer loop for convergence, respectively. The computational complexities of $\textbf{Algorithm~2}$ and $\textbf{Algorithm~3}$ can be expressed as $\mathcal{O}\left(MK\log\left(MK\right)\right)$ and $\mathcal{O}(M^2KT_{\text{swap}})$, where $T_{\text{swap}}$ denotes the number of swaps in the matching algorithm. Therefore, the overall computational complexity of $\textbf{Algorithm~4}$ can be given by $O\left(MK\log\left(MK\right)+M^2KT_{\text{swap}}+I_{\mathrm{in}}I_{\mathrm{out}}\left(\left(I_{\text{iter,1}}M^{3.5}\right.\right.\right.$  $+I_{\text{iter,2}}\left(KN\right)^{1.5}+MKN^{3}+I_{\text{iter,4}}MKN\Big{)}\Big{)}\Big{)}$.

\section{Simulation Results}
In this section, numerical assessments are conducted to evaluate the effectiveness and superiority of our proposed Matching-PDD algorithm in the WD-PASS-based NOMA framework. 
\vspace{-0.4cm}
\subsection{Simulation Setup}
In simulations, we consider a WD-PASS-based NOMA scenario, where the BS is equipped with three waveguides, i.e., $K=3$. The coordinates of the feed points on the three waveguides are $\left(0,-\frac{W}{2},d\right)$, $\left(0,0,d\right)$, and $\left(0,\frac{W}{2},d\right)$ meters, respectively.  $M$ users are randomly distributed within a rectangular area $\mathcal{S}=\Big{\{}\left[x,y,0\right]\big{|}x\in(0,L),y\in\left(-\frac{KW}{2},\frac{KW}{2}\right)\Big{\}}$. For initialization, we assume that PAs positions are uniformly spaced along each waveguide, and the BS transmit power is allocated uniformly among all $M$ users. Unless otherwise specified, the system parameter values are presented in Table~\ref{table:parameters}. 

\begin{table}[h!] 
\caption{Simulation parameters}
\centering
\label{table:parameters}
\begin{tabular}{|m{1.4cm}|p{4.5cm}|m{1.4cm}|}
\hline
Parameter & Description & Value\\
\hline
$f$ & Carrier frequency & $6$~GHz \\ \hline
{$\Delta$} & Minimum PAs distance  & $\frac{\lambda}{2}$\\ \hline
$R_{\text{min}}$ & QoS requirement of each user & $0.05$~bps/Hz\\ \hline
$n_{\text{eff}}$ & Effective refractive index of a dielectric waveguide & $1.4$\\ \hline
$\sigma^2$ & Noise power  & $-90$~dBm\\ \hline
$P_{\text{max}}$ & Maximum BS transmit power & $15$~dBm\\ \hline
$W$ & Distance between two waveguides & $10$~m\\ \hline
$L$ & Length of each waveguide & $10$~m\\ \hline
$N$ & Number of PAs on each waveguide & $8$\\ \hline
$d$ & Height of all waveguides & $3$~m\\ \hline
$\bar{q}$ & Maximum number of the users served by each waveguide & $3$ \\ \hline
$\epsilon_1$ & Convergence threshold of the inner loop in \textbf{Algorithm~2} & $10^{-4}$ \\ \hline
$\epsilon_2$ & Convergence threshold of the outer loop in \textbf{Algorithm~2} & $10^{-5}$ \\ \hline
$\rho$ & Initial Lagrangian penalty factor & $10^{-4}$ \\ \hline
\end{tabular}
\end{table}
\subsection{Baseline Schemes}
\begin{itemize}
    \item \textbf{Conventional antenna systems}: In this case, we consider a conventional BS, which is equipped with a uniform linear array (ULA) with $K$ fixed-position antennas. The ULA is located parallel to the $y$-axis, and the spacing between two antennas is given by $\frac{\lambda}{2}$. We assume that the BS employs fully-digital beamforming, while the user scheduling is designed by exploiting \textbf{Algorithm~\ref{alg:user_scheduling}}.
    \item \textbf{Random PA placement scheme (RPP)}: In this case, we consider random PAs positions, where the power allocation and user scheduling are optimized by exploiting  \textbf{Algorithm~\ref{alg:pdd}} and \textbf{Algorithm~\ref{alg:user_scheduling}}, respectively.     
    \item \textbf{WD-PASS-OMA scheme}: In this case, the users in each cluster are accessed through TDMA, where the joint power allocation, pinching beamforming, and user scheduling problem is optimized via \textbf{Algorithm~\ref{alg:matching-pdd}}.
    \item \textbf{Exhaustive search-PDD scheme} : In this case, we consider an exhaustive search method to design user scheduling by searching all possible user-to-waveguide assignments, while \textbf{Algorithm~\ref{alg:pdd}} is utilized to optimize power allocation and pinching beamforming.
\end{itemize}

\subsection{Convergence of the Proposed Algorithms}
 \begin{figure}
    \centering    \includegraphics[width=1.00\linewidth,height=0.8\linewidth]{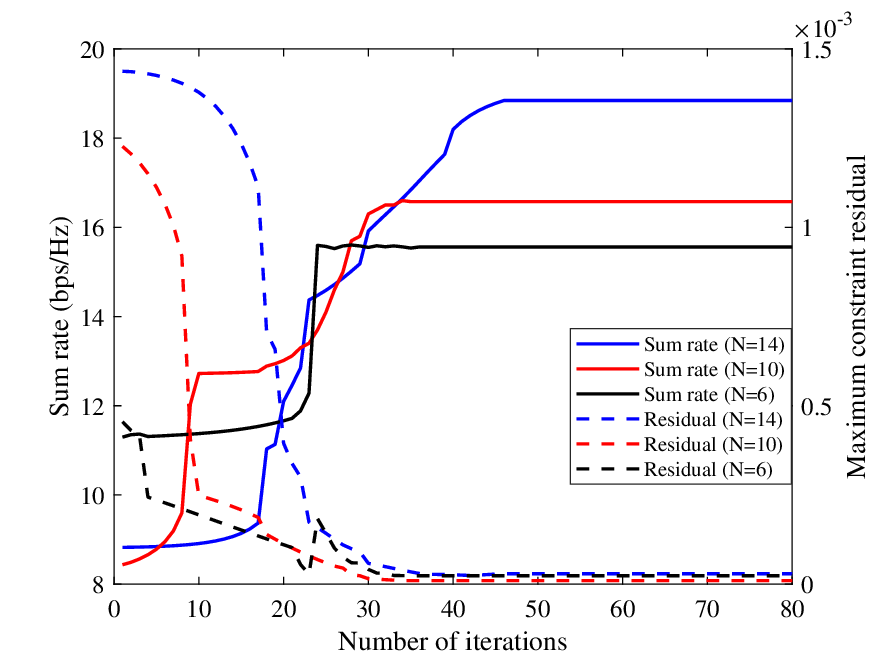}
    \caption{Convergence behaviors of the PDD algorithm.}
    \label{fig:sum-rate-iter}
\end{figure}
In Fig.~\ref{fig:sum-rate-iter}, we first evaluate the convergence of the proposed PDD algorithm in \textbf{Algorithm~\ref{alg:pdd}} versus the number of PAs, i.e, $N$, where we set $M=6$. To comprehensively demonstrate the performance of the PDD algorithm, we plot both the sum rate and the maximum constraint residual, i.e., $\|\mathbf{B}\|_{\infty}$. It can be observed that, the PDD algorithm simultaneously increases the system sum rate and reduces constraint residuals with the number of iterations. It is also shown that, the sum rate increases slowly at the beginning of the iterations, when the residuals of equality constraints dominate the objective function. When the constraint residuals become small, the system sum rate increases significantly. Specifically, the maximum constraint residual $\|\mathbf{B}\|_{\infty}$ decreases rapidly within $30$ iterations, and eventually converges towards the predefined threshold $\epsilon_2$, which demonstrates that the PDD algorithm can guarantee the satisfactory of the equality constraints~\eqref{eq:b_u}, ~\eqref{eq:b_theta}, and~\eqref{eq:b_q}. Moreover, we can also find that the number of iterations for the convergence increases with the increment of $N$. This is because, the increment in the number of PAs positions leads to higher computational complexity, which consequently results in a slower convergence speed for the proposed PDD algorithm.

\subsection{Sum Rate Versus Transmit Power}
 \begin{figure}
    \centering    \includegraphics[width=1.00\linewidth,height=0.8\linewidth]{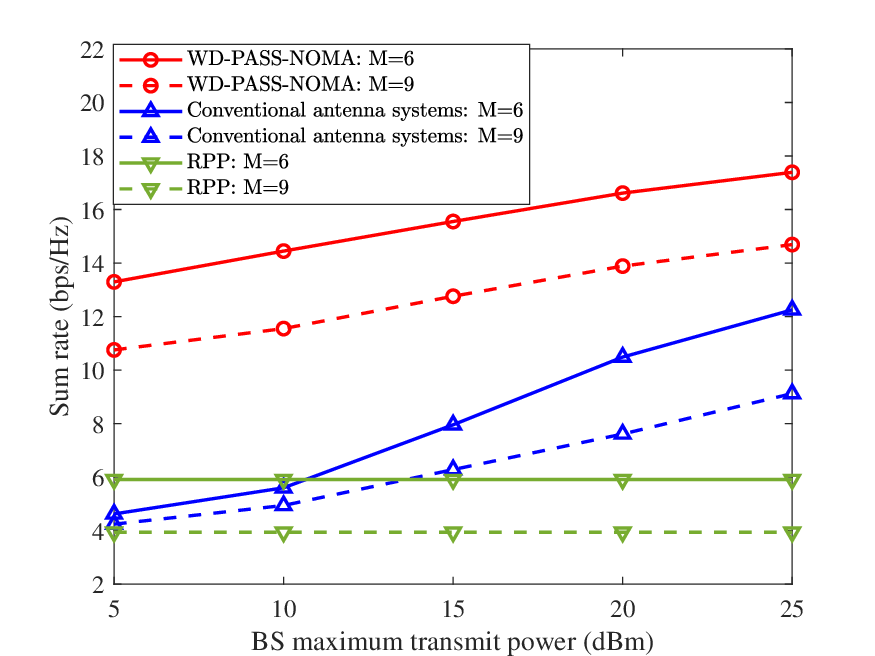}
    \caption{Sum rate versus maximum transmit power.}
    \label{fig:p}
\end{figure}

In Fig.~\ref{fig:p}, we present the sum rate of different schemes versus maximum BS transmit power $P_\text{max}$. As can be observed, the proposed WD-PASS-NOMA scheme achieves significant performance gain over conventional antenna systems. This is because, on the one hand, by enabling dynamic PA deployment proximate to target users, PASS achieve a stronger LoS link, overcoming the high path loss inherent in centralized deployments. On the other hand, since multiple PAs can be activated on a single waveguide at no additional cost, increasing the number of PAs in the PASS architecture provides more degrees of freedom (DoFs) for pinching beamforming. Moreover, it can also be observed that, the proposed WD-PASS-NOMA scheme achieves superior performance than the RPP benchmarks,  demonstrating the critical role of pinching beamforming in enhancing system performance. In addition, when the number of users $M$ increases, we observe that the sum rate decreases for all schemes. The reason is that the BS needs to guarantee QoS constraints for more users in a communication scenario with more severe interference, thereby reducing the overall sum rate.

\subsection{Sum Rate Versus Waveguide Length}
 \begin{figure}
    \centering    \includegraphics[width=1.0\linewidth,height=0.8\linewidth]{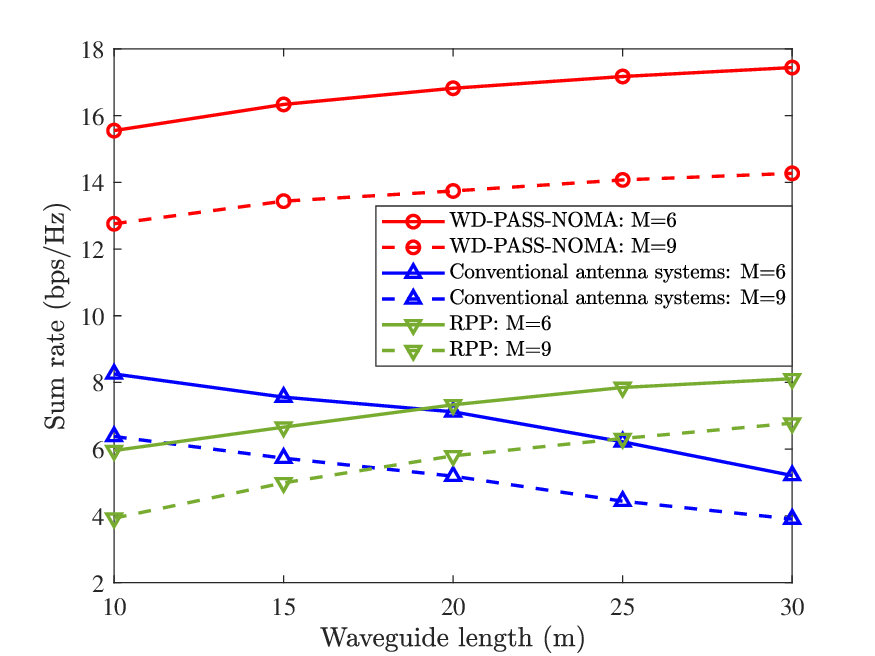}
    \caption{Sum rate versus waveguide length.}
    \label{fig:sum-rate-L}
\end{figure}

In Fig.~\ref{fig:sum-rate-L}, we investigate the achievable sum rate versus the waveguide length, i.e., $L$.  As can be observed, the sum rate achieved by the conventional antenna system decreases with the increment of $L$. The reason is that, the path loss between users and PAs increases with the increment of the waveguide length. Meanwhile, it is interesting to observe that, the system sum rates achieved by the WD-PASS-NOMA and the RPP schemes increase when the $L$ becomes larger. This is becuse, on the one hand, the pinching beamforming can address the increasing path loss by intelligently adjusting PA positions to establish strong LoS links near the users. On the other hand, the intra-cluster and inter-cluster user interference will gradually decrease as the distribution range of users expands.

\subsection{Sum Rate Versus Number of PAs}
 \begin{figure}
    \centering    \includegraphics[width=1.00\linewidth,height=0.8\linewidth]{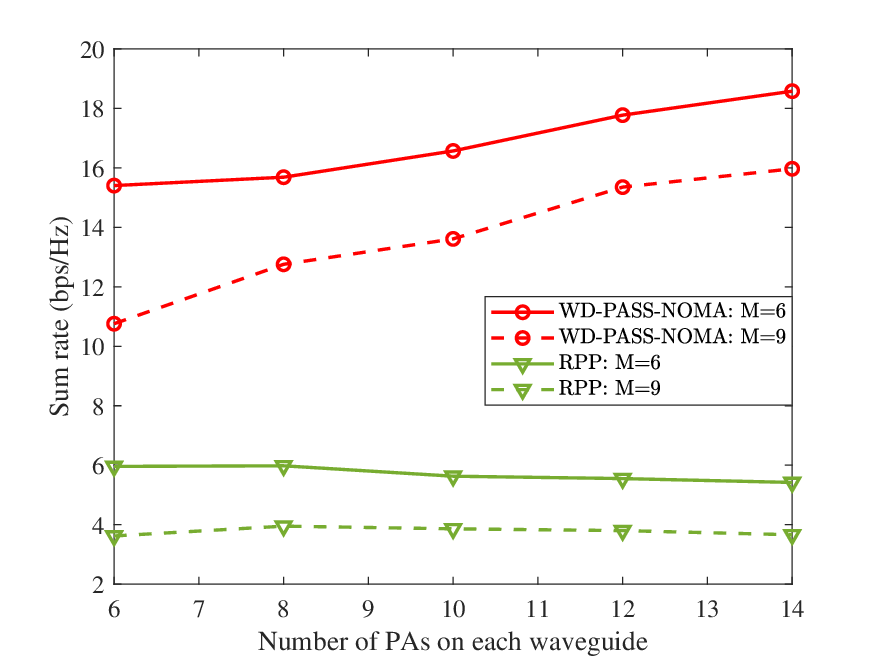}
    \caption{Sum rate versus number of PAs.}
    \label{fig:sum-rate-N}
\end{figure}

Fig.~\ref{fig:sum-rate-N} depicts the sum rate versus the number of PAs on each waveguide, i.e., $N$. As can be observed, the sum rate achieved by the WD-PASS-NOMA scheme increases with the increment of $N$. This is because, a larger number of available PAs provides higher DoFs for the pinching beamforming, thereby allowing for more effective intended signal enhancement and interference mitigation. In contrast, the performance of the RPP scheme remains nearly constant and even slightly degrades as $N$ increases. This indicates that simply increasing the number of antennas without intelligent PAs placement fails to effectively leverage the spatial resources. The obtained results demonstrate the superiority of the proposed WD-PASS-NOMA framework for properly adjusting PAs positions to improve the pinching beamforming performance.

 \begin{figure}
    \centering    \includegraphics[width=1.00\linewidth,height=0.8\linewidth]{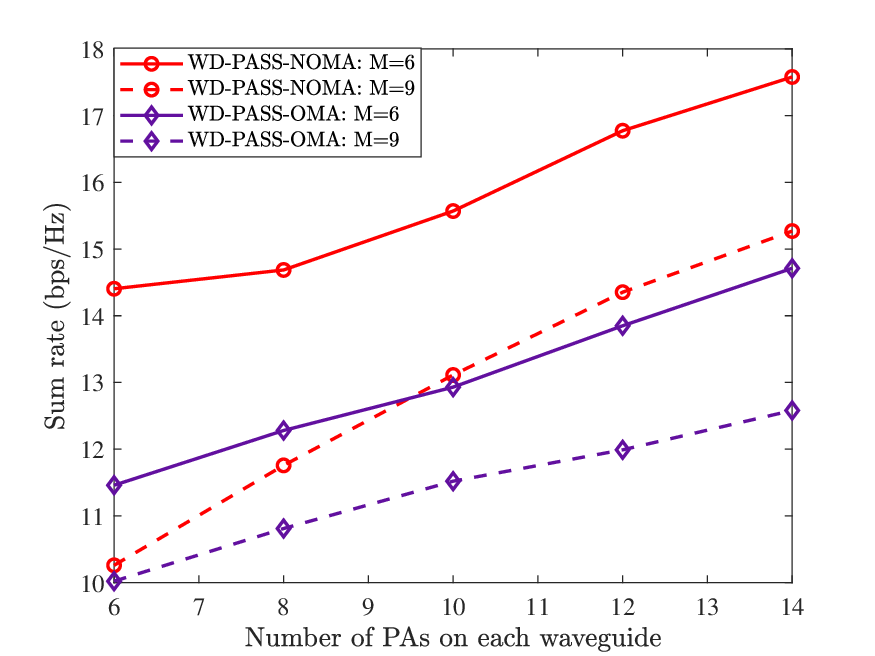}
    \caption{Sum rate versus number of PAs.}
    \label{fig:sum-rate-OMA}
\end{figure}
\subsection{Performance Comparison Between NOMA and OMA Schemes in PASS}
To further evaluate the efficiency of the proposed WD-PASS-NOMA scheme, we compare it against the WD-PASS-OMA scheme with $P_{\text{max}}=10~\text{dBm}$ in Fig.~\ref{fig:sum-rate-OMA}. It can be observed that the sum rate achieved by the WD-PASS-OMA scheme is much lower than that achieved by the WD-PASS-NOMA scheme. For instance, the sum rate achieved by the proposed WD-PASS-NOMA scheme is around $17.8~\text{bps/Hz}$ when we set $N=14$ and $M=6$, while that achieved by the WD-PASS-OMA scheme is around $15.8~\text{bps/Hz}$. The main reason is that NOMA exploits additional DoFs by power-domain multiplexing to improve its efficiency. The NOMA scheme intentionally allows users to share all frequency resources non-orthogonally, and manages the severe interference by leveraging transmit power allocation at the BS and SIC at the users, which overcomes the orthogonal resource allocation limit of the OMA system. These observations indicate that the application of NOMA into PASS can further improve the spectral efficiency.

 \begin{figure}
    \centering    \includegraphics[width=1.00\linewidth,height=0.8\linewidth]{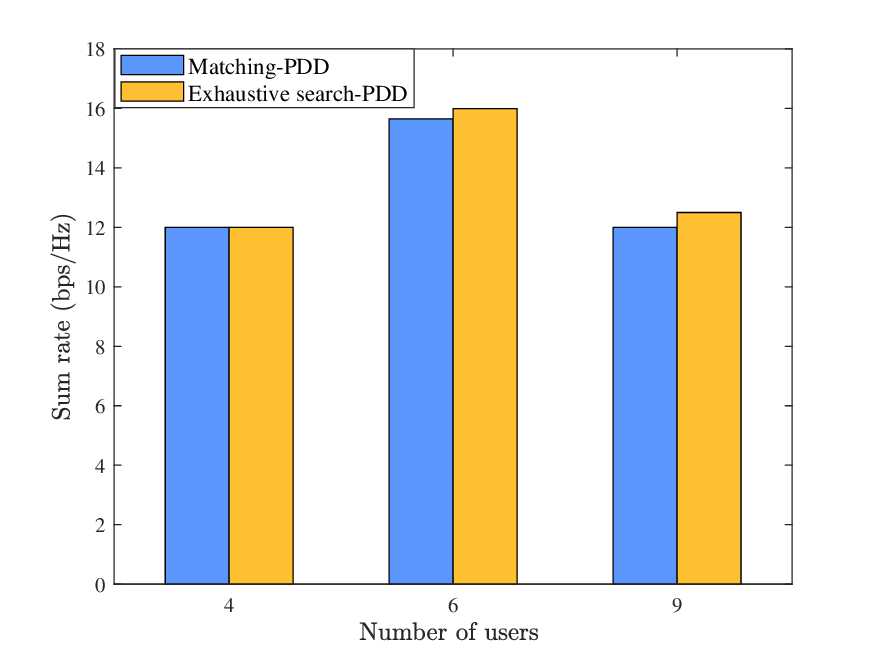}
    \caption{Sum rate versus number of users.}
    \label{fig:sum-rate-matching}
\end{figure}
\subsection{Performance Comparison Between Matching and Exhaustive Search Schemes}

In Fig.~\ref{fig:sum-rate-matching}, we demonstrate the effectiveness of the proposed Matching-PDD algorithm by comparing it with the Exhaustive search-PDD scheme under cases with different number of users. It can be first observed that, when the number of users is set as $4$, the Matching-PDD algorithm achieves the same performance as the Exhaustive search-PDD algorithm. This is because, the potential matching pairs between waveguides and users is limited when the number of users is small, which makes it smooth for the matching algorithm to find the optimal match state.  As the number of users increases, the Matching-PDD algorithm still shows near-optimal performance, with the computational complexity much lower than that of the exhaustive search, as demonstrated in Section III-C. For example, the sum rate of the Matching-PDD algorithm achieve around 97$\%$ and 95$\%$ to that of the Exhaustive search-PDD scheme for $m=6$ and $m=9$, respectively. These observations demonstrate the effectiveness of the proposed algorithm for solving the user scheduling problem in the WD-PASS-based NOMA framework. 

\section{Conclusions}
This paper proposed a novel WD-PASS-based NOMA framework for multi-user downlink communications, where each NOMA user cluster is served by one dedicated waveguide. To address the non-convex problem of jointly optimizing PAs positions, power allocation, and user scheduling, an efficient two-step framework that decomposed the original formulation into two tractable problems was proposed. Specifically, a PDD algorithm based on the BCD method was developed to jointly optimize power allocation and pinching beamforming. Subsequently, the user scheduling problem was reformulated as a many-to-one matching problem and solved by a low-complexity matching-based algorithm. Simulation results demonstrated that the proposed WD-PASS-based NOMA framework achieved significant sum-rate gains over conventional fixed-position antenna systems and the WD-PASS-based OMA system, which demonstrated the effectiveness of the integration of WD-PASS with the NOMA protocol for serving multiple users in reality.

\bibliographystyle{IEEEtran} 
\bibliography{sample}

\end{document}